# Systematic determination of coupling constants in spin clusters from broken-symmetry mean-field solutions

Shadan Ghassemi Tabrizi

*Department of Chemistry, University of Potsdam, Karl-Liebknecht-Str. 24-25, D-14476, Potsdam-Golm, Germany*

shadan.ghassemi@uni-potsdam.de

**Abstract.** Quantum-chemical calculations aimed at deriving magnetic coupling constants in exchange-coupled spin clusters commonly utilize a broken-symmetry (BS) approach. This involves calculating several distinct collinear spin configurations, predominantly by density-functional theory (DFT). The energies of these configurations are interpreted in terms of the Heisenberg model, $\underline{H} = \sum_{i<j} J_{ij} \underline{\mathbf{s}}_i \cdot \underline{\mathbf{s}}_j$, to determine coupling constants $J_{ij}$ for spin pairs. However, this energy-based procedure has inherent limitations, primarily in its inability to provide information on isotropic spin interactions beyond those included in the Heisenberg model. Biquadratic exchange or multi-center terms, for example, are usually inaccessible and hence assumed to be negligible. The present work introduces a novel approach employing BS mean-field solutions, specifically Hartree–Fock wave functions, for the construction of effective spin Hamiltonians. This expanded method facilitates the extraction of a broader range of coupling parameters by considering not only the energies, but also Hamiltonian and overlap elements between different BS states. We demonstrate how comprehensive $s = \frac{1}{2}$ Hamiltonians, including multi-center terms, can be straightforwardly constructed from a complete set of BS solutions. The approach is exemplified for small clusters within the context of the half-filled single-band Hubbard model. This allows to contrast the current strategy against exact results, thereby offering an enriched understanding of the spin-Hamiltonian construction from BS solutions.

## 1. Introduction

Effective spin models are an indispensable tool in the analysis of extended magnetic solids[1] as well as finite exchange-coupled clusters,[2] including molecular magnets.[3,4] The isotropic Heisenberg model is often of primary interest. It describes exchange interactions between open-shell sites in terms of pairwise spin couplings. The respective coupling constants serve as the model's free parameters and offer insight into the nature and strength of the magnetic interactions. The Heisenberg model forms the foundation for the calculation of thermodynamic properties, such as magnetic susceptibilities, heat capacities, or phase transitions. Furthermore, it aids in the interpretation of inelastic neutron scattering[5,6] (INS) spectra and in elucidating various other spectroscopic features.

Quantum chemistry offers a path to compute Heisenberg coupling constants. Such calculations provide microscopic insights into the origins of magnetic behavior, and thus carry the potential to aid in the design of novel systems with desired characteristics. For the determination of exchange parameters from first principles, multi-configurational wave-function approaches like the complete active space self-consistent field (CASSCF) method, can be employed. By coupling CASSCF with a Density Matrix Renormalization Group (DMRG) solver,[7,8] the active space of the calculations can be significantly expanded, and methods like CASPT2[9] (CAS with second-order perturbation theory) or NEVPT2[10] (N-electron valence-state PT2) incorporate dynamic correlation effects into CASSCF. Despite these advancements, accurately calculating exchange couplings, even for dimers, remains challenging.[11–14] Indeed, as the number of interacting spins increases, the strong-correlation problem becomes more complex, leading to a wide array of electronic levels with various total spins due to effective interactions between local spins. This renders the task of approximating all these states and establishing a comprehensive and accurate set of coupling constants intractable. In this regard, it is also worth mentioning a strategy by Mayhall and Head-Gordon[15,16] which significantly reduces the number of essential configurations as it only targets eigenstates in the sectors with maximal total spin $S_{max}$ and $S_{max} - 1$. However, due to the limited number of levels, assumptions about the spin-Hamiltonian parameters are necessary. In addition, CASSCF calculations may reveal significant discrepancies when comparing $J$ values from the energy differences of $S_{max}$ and $S_{max} - 1$ levels with those obtained from states of the lowest multiplicity.[17] Answering the question whether deviations from a Landé-type level structure in dimers should be primarily attributed to higher-order exchange terms such as biquadratic exchange,[14] or to a dependence of the accuracy of a theoretical method on the multiplicities of distinct spin levels,[17] requires a reliable treatment of static and dynamic correlation.

Broken-symmetry (BS) methods, particularly BS-DFT, are a prominent alternative to the intrinsic computational challenges of *ab initio* calculations of spin eigenstates in exchange-coupled clusters.[18–24] Instead of approximating eigenstates, BS methods focus on specific spin configurations, which break spin symmetry and (where applicable) point-group (PG) symmetry. The energies of BS states can be used to estimate exchange couplings. This energy-based mapping onto spin Hamiltonians represents a considerable advantage in terms of computational efficiency and scalability and contributes to understanding magnetic properties in larger spin clusters. For example, BS-DFT was applied to single-molecule magnets such as $Mn_{12}$,[25–27] and even to $Mn_{70}$ or $Mn_{84}$ rings.[28] Such investigations would be infeasible with methods that attempt to find approximate solutions to the electronic Schrödinger equation.

However, the basic BS approach has certain limitations. Due to the restricted number of different spin configurations, the energy-based mapping might not yield a comprehensive set of parameters required for capturing some intricate features.[29–31] An example where the standard mapping falls short of determining all parameters is a system composed of four $s = \frac{1}{2}$ sites.[29] If PG symmetry is absent (or if the sites are arranged symmetrically on a square), there are six (two) different pairwise exchanges and three (two) independent four-center terms, adding up to nine (four) parameters in total. However, it is only possible to calculate eight (four) unique spin configurations and to extract a maximum of seven (three) spin-Hamiltonian parameters, see Results section. This discrepancy showcases a limitation of the conventional approach. As a result, in specific cases, the models generated might fall short of fully capturing the magnetic behavior.

Furthermore, the recent adoption of coupled-cluster theory for calculations of spin couplings[17] highlights challenges of BS approaches. Computational methods that implicitly describe some strong-correlation effects, like coupled cluster at sufficiently high order, give rise to BS solutions that resemble exact eigenstates of the system.[17] While there are established strategies that can adjust for deviations between BS solutions and ideal mutually orthogonal spin configurations,[32,33] particularly evident in strongly antiferromagnetic situations, a more systematic process is lacking. The present work therefore aims to develop a robust generalized framework for the construction of spin Hamiltonians from BS solutions. A central idea is to devise an effective Hamiltonian for $s = \frac{1}{2}$ systems using a full set of BS solutions. The resulting energy levels essentially correspond to solving a non-orthogonal configuration interaction (NOCI) problem within the space spanned by the BS solutions.

We first briefly review the utilization of BS solutions in the calculation of spin Hamiltonians, along with an exposition of their limitations. This is followed by the formulation of a broader theoretical groundwork for constructing $s = \frac{1}{2}$ Hamiltonians using a comprehensive set of BS solutions. Our approach draws upon concepts from canonical effective-Hamiltonian theory.[34–36] Subsequently, various types of spin interactions are addressed, with a particular emphasis on multi-center terms. The primary aim of this paper is to introduce a novel method for constructing spin Hamiltonians, rather than deepen the understanding of magnetic interactions in any specific system. The Results section therefore showcases applications to small clusters within the single-band Hubbard model. This model serves as a simplistic testbed for the proposed methodology, allowing for comparisons with exact-diagonalization results. In closing, future directions are briefly discussed, such as extensions of the present methodology to systems with higher spins or to realistic molecular exchange-coupled clusters.

## 2. Theory

This section starts with a summary of the canonical effective-Hamiltonian theory developed by Bloch[34] and refined by des Cloizeaux.[35] For a detailed review including applications, the reader may refer to Ref. 36. This establishes the foundation for a new strategy to derive a comprehensive set of spin-Hamiltonian parameters for $s = \frac{1}{2}$ systems, from both energies and off-diagonal matrix elements between different BS solutions. We subsequently provide a reminder on the spin Hamiltonian's emergence within the half-filled Hubbard model in the strong-coupling limit and highlight the onset of multi-center exchange, which we then briefly explore in broader terms. The role of PG symmetry in reducing the count of independent parameters is addressed.

**Effective-Hamiltonian theory.** Suppose that a set of low-energy eigenstates $|\Psi_i\rangle$, $i = 1, 2, ..., D$, of the electronic Hamiltonian, $\hat{H}|\Psi_i\rangle = E_i|\Psi_i\rangle$, has been calculated (exactly or approximately), which constitute the target space for an effective Hamiltonian (the spin Hamiltonian). For example, for $N$ sites with $s = \frac{1}{2}$, the dimension of the target space is $D = 2^N$. One needs to establish a projector $P_0$ for mapping the target space onto the model space.[37] The definition of $P_0$ implies a one-to-one relationship between electronic states and effective (spin) states, which is usually ambiguous.[38,39] However, for the single-band Hubbard model at half filling or for charge-neutral clusters of minimal-basis H atoms, $P_0$ can be uniquely defined. For a dimer, the simplest example, $P_0$ is given in Eq. (1),

$$P_0 = |\uparrow\uparrow\rangle\langle\Uparrow\Uparrow| + |\uparrow\downarrow\rangle\langle\Uparrow\Downarrow| + |\downarrow\uparrow\rangle\langle\Downarrow\Uparrow| + |\downarrow\downarrow\rangle\langle\Downarrow\Downarrow| \,, \quad (1)$$

where $|\Uparrow\Uparrow\rangle = \hat{c}^\dagger_{1\uparrow}\hat{c}^\dagger_{2\uparrow}|0\rangle$ is an electronic state ($|0\rangle$ is the vacuum) hosting a spin-up electron on each site, and $|\uparrow\uparrow\rangle$ is a spin configuration residing in the space of the effective spin model, with local $z$-projections of $m_1 = m_2 = +\tfrac{1}{2}$, and so forth.

Model-space projections $|\bar{\Psi}_i\rangle$ of orthonormal electronic eigenstates $|\Psi_i\rangle$,

$$|\bar{\Psi}_i\rangle = P_0 |\Psi_i\rangle \,. \quad (2)$$

undergo symmetrical (least moving) orthogonalization,

$$|\tilde{\Psi}_i\rangle = \sum_j (\mathbf{S}^{-1/2})^*_{ij} |\bar{\Psi}_j\rangle \,, \quad (3)$$

where $\bar{S}_{ij} \equiv \langle\bar{\Psi}_i|\bar{\Psi}_j\rangle$. An effective Hamiltonian $\underset{\sim}{H}$ is then constructed, Eq. (4),

$$\underset{\sim}{H} \equiv \sum_i E_i |\tilde{\Psi}_i\rangle\langle\tilde{\Psi}_i| \,, \quad (4)$$

which exactly reproduces the spectrum of electronic energies $E_i$ of the chosen states.

When spin symmetry is broken, for instance, by including spin-orbit coupling into the electronic Hamiltonian, the spin Hamiltonian $\underset{\sim}{H}$ will comprise both isotropic and anisotropic contributions.[2,13,40] Even when spin symmetry is conserved, $\underset{\sim}{H}$ may include many terms in addition to pairwise Heisenberg-type interactions (see below). Except for the simplest cases that only allow a small number of parameters, a systematical formulation using irreducible tensor operators (ITOs[2,41]) is advised for expanding $\underset{\sim}{H}$ and thereby defining the independent model parameters. Trace formulas for ITO expansions of spin Hamiltonians were introduced in Ref. 40.

**Effective-Hamiltonian theory based on BS solutions.** We propose a novel procedure, intuitively grounded in canonical effective-Hamiltonian theory, to map a comprehensive isotropic $s = \tfrac{1}{2}$ Hamiltonian from BS solutions. Instead of approximating $D = 2^N$ eigenstates $|\Psi_i\rangle$, a challenging task within the strong-correlation context, a respective number of (broken-symmetry) mean-field states $|\Phi_i\rangle$ is converged. The spin densities at the open-shell sites of each $|\Phi_i\rangle$ correspond to a specific spin configuration. For instance, in a dimer, $|\Phi_1\rangle = |\underline{\Uparrow\Uparrow}\rangle$, $|\Phi_2\rangle = |\underline{\Downarrow\Uparrow}\rangle$, $|\Phi_3\rangle = |\underline{\Uparrow\Downarrow}\rangle$, and $|\Phi_4\rangle = |\underline{\Downarrow\Downarrow}\rangle$. The underscore notation is employed to differentiate mean-field states from the previously introduced model states. In a minimal basis $|\underline{\Uparrow\Uparrow}\rangle = |\Uparrow\Uparrow\rangle$, yet $|\underline{\Uparrow\Downarrow}\rangle \neq |\Uparrow\Downarrow\rangle$. High-spin (HS) states $|\Phi_1\rangle$, $|\Phi_4\rangle$ and BS states $|\Phi_2\rangle$, $|\Phi_3\rangle$ are pairwise related by a spin flip. If calculated by an unrestricted method like UHF, $|\Phi_i\rangle$ breaks spin symmetry.[42] The distinct mean-field solutions $|\Phi_i\rangle$ have non-zero overlap, unless differing in spin projection $M$ or falling into separate PG species, and are hence orthogonalized, as shown in Eq. (5),

$$|\Xi_i\rangle = \sum_j (\mathbf{S}^{-1/2})^*_{ij} |\Phi_j\rangle ,\qquad(5)$$

where $S_{ij} \equiv \langle \Phi_i | \Phi_j \rangle$. The $|\Xi_i\rangle$ span the target space and serve in constructing the effective Hamiltonian, echoing the canonical procedure based on eigenstates $|\Psi_i\rangle$. Consequently, the target space is projected onto the model space, Eq. (6),

$$|\bar{\Xi}_i\rangle = P_0 |\Xi_i\rangle .\qquad(6)$$

Unlike canonical effective-Hamiltonian theory, which only requires consistently chosen phases of model-space states, an additional necessity is ensuring that states $|\bar{\Xi}_i\rangle$ have consistent relative phases. Specifically, the relative sign of BS solutions $|\Xi_i\rangle$ (considering exclusively real wave functions here) should be selected in accord with the corresponding model-space states. In the Hubbard model (see next subsection), the overlap between $|\Xi_i\rangle$ and its corresponding model-space state approaches 1 in the limit $\frac{t}{U} \to 0$.

Once signs are set, the $|\bar{\Xi}_i\rangle$ are orthogonalized, as in Eq. (7),

$$|\tilde{\bar{\Xi}}_i\rangle = \sum_j (\bar{\mathbf{S}}^{-1/2})^*_{ij} |\bar{\Xi}_j\rangle ,\qquad(7)$$

where $\bar{S}_{ij} \equiv \langle \bar{\Xi}_i | \bar{\Xi}_j \rangle$. We are only dealing with real UHF solutions $|\Phi_i\rangle$, rendering the complex-conjugation symbol unnecessary. Finally, the effective Hamiltonian is derived

following Eq. (8), employing orthogonal projections $|\tilde{\Xi}_i\rangle$ and Hamiltonian matrix elements $H_{ij} = \langle \Xi_i | \hat{H} | \Xi_j \rangle$ between orthogonal unprojected states,

$$\utilde{H}_{BS} \equiv \sum_{i,j} H_{ij} |\tilde{\Xi}_i\rangle \langle \tilde{\Xi}_j | . \quad (8)$$

The calculation of matrix elements $S_{ij}$ and $H_{ij}$ between non-orthogonal Slater determinants can be efficiently accomplished using single-particle transition-density matrices based on the extended Wick's theorem. A detailed derivation can be found in Ref. 43, and recent applications include Refs. 44,45. We however did not utilize this powerful formalism, but instead worked in the full configuration interaction (FCI) basis; this would not be feasible for significantly larger systems.

In this study, all mean-field solutions are of UHF type, thus conserving the $z$-component of spin. Therefore, $\utilde{H}_{BS}$ possesses $\utilde{S}_z$ symmetry, $[\utilde{H}_{BS}, \utilde{S}_z] = 0$. However, the implemented procedure does not generally produce an isotropic $\utilde{H}_{BS}$, implying, $[\utilde{H}_{BS}, \utilde{S}_x] \neq 0$, $[\utilde{H}_{BS}, \utilde{S}_y] \neq 0$, even when $\hat{H}$ preserves spin symmetry. $\utilde{\mathbf{S}}^2$ symmetry can be easily restored by summing the projections of $\utilde{H}_{BS}$ onto all total-spin sectors,

$$\hat{\utilde{H}}_{BS} = \sum_{S=S_{\min}}^{S_{\max}} \utilde{P}_S \utilde{H}_{BS} \utilde{P}_S , \quad (9)$$

utilizing Löwdin's spin projector, Eq. (10),

$$\utilde{P}_S = \prod_{l \neq S} \frac{\utilde{\mathbf{S}}^2 - l(l+1)}{[S(S+1) - l(l+1)]} . \quad (10)$$

Given $[\hat{\utilde{H}}_{BS}, \utilde{\mathbf{S}}^2] = 0$, each level of $\hat{\utilde{H}}_{BS}$ possesses a well-defined total spin. Nonetheless, full spin-rotational symmetry is not achieved, $[\hat{\utilde{H}}_{BS}, \utilde{S}_x] \neq 0$, $[\hat{\utilde{H}}_{BS}, \utilde{S}_y] \neq 0$, leading to residual anisotropic contributions. Consequently, magnetic substates forming a spin multiplet exhibit (slightly) differing energies in distinct $|M|$ sectors. This artificial anisotropy will be quantified for an example in the Results section. A pragmatic approach to handling "anisotropy" would be to only consider the sector with the smallest $|M|$, namely $M = 0$ or $M = \frac{1}{2}$, which encompasses all spin levels, or by tracing over all $M$ sectors, which corresponds to a weighted average. We do not attempt full restoration of spin-rotational symmetry in this work.

On the other hand, $\underset{\sim}{H}_{BS}$ conserves PG symmetry. By employing PG operators on a single UHF solution, one can obtain symmetry-equivalent degenerate solutions. Additionally, the basis includes degenerate determinants with all spins flipped, and all orbitals have real-valued coefficients. As such, $\underset{\sim}{H}_{BS}$ also maintains invariance under time reversal as well as complex-conjugation symmetry.

**Hubbard model.** The Hubbard model is one of the simplest models of interacting electrons in a tight-binding lattice. It consists of two main terms: an on-site interaction term representing the repulsion between electrons occupying the same site (on-site repulsion $U \geq 0$) and a hopping term describing the movement of electrons between neighboring sites (hopping integral $t \geq 0$),

$$\hat{H} = U \sum_{i=1}^{N} \hat{n}_{i\uparrow}\hat{n}_{i\downarrow} - t \sum_{\langle i,j \rangle} \hat{c}_{i\sigma}^{\dagger}\hat{c}_{j\sigma} \ , \tag{11}$$

where $\hat{c}_{i\sigma}^{\dagger}$ ($\hat{c}_{i\sigma}$) creates (annihilates) an electron with spin $\sigma = \uparrow, \downarrow$ at site $i$, and $\hat{n}_{i\sigma} = \hat{c}_{i\sigma}^{\dagger}\hat{c}_{i\sigma}$ is the local occupation number. The second summation runs over all unique pairs $\langle i,j \rangle$ of neighboring sites. At half-filling, the number of electrons equals the number $N$ of sites, and the Hubbard model can be reduced to the Heisenberg model in the strong-coupling limit $U \gg t$. In this limit, the system is effectively in a Mott insulating state with one electron per site, and the low-energy physics is dominated by spin interactions rather than charge movements. The reduction can be done by a canonical transformation,[46] which involves expanding the Hubbard Hamiltonian in powers of $\frac{t}{U}$, treating the hopping term as a perturbation around the limit of infinite on-site repulsion. At the leading second order, this procedure results in an effective spin Hamiltonian which is just the Heisenberg model, with an antiferromagnetic exchange interaction $J = 4t^2/U$. The exchange term arises from virtual processes in which an electron hops to a neighboring site already occupied by an electron of opposite spin (costing an energy $U$), and then hops back. The microscopic mechanisms responsible for magnetic interactions in real materials are far more complex and can lead to both ferromagnetic and antiferromagnetic interactions. In the perturbation expansion of the Hubbard model, four-center (4c) terms start to appear in the fourth order of perturbation theory (PT4), while 6c terms and other higher-order multi-center terms occur from the sixth order and beyond. The contribution of multi-exchange terms to observable properties and spectra tends to be small. However, specific symmetries or geometries of the system can enhance these interactions.[47,48]

**Multi-exchange terms.** A complete set of linearly independent multi-center terms can be systematically generated by coupling irreducible tensor-operators (ITOs) successively. Local rank-1 ITOs serve as building blocks, with spherical components $\underline{T}^{(1)}_{+1}(\mathbf{s}_i) = -\frac{1}{\sqrt{2}}(\underline{s}_{i,x} + i\underline{s}_{i,y})$, $\underline{T}^{(1)}_{0}(\mathbf{s}_i) = \underline{s}_{i,z}$ and $\underline{T}^{(1)}_{-1}(\mathbf{s}_i) = \frac{1}{\sqrt{2}}(\underline{s}_{i,x} - i\underline{s}_{i,y})$ forming a rank-1 ITO $\underline{\mathbf{T}}^{(1)}(\mathbf{s}_i)$. The coupling of two ITOs $\underline{\mathbf{T}}^{(k_i)}$ and $\underline{\mathbf{T}}^{(k_j)}$, using Wigner-3$j$ symbols, generates a compound ITO $\underline{\mathbf{T}}^{k_i k_j (k_{ij})}$ of rank $k_{ij}$ with components $\underline{T}^{k_i k_j (k_{ij})}_q$ ($q = -k_{ij}, -k_{ij}+1, \ldots, +k_{ij}$),

$$T^{k_i k_j (k_{ij})}_{\sim q}(\mathbf{s}_i, \mathbf{s}_j) \equiv \left[\underline{\mathbf{T}}^{(k_i)}(\mathbf{s}_i) \otimes \underline{\mathbf{T}}^{(k_j)}(\mathbf{s}_j)\right]^{(k_{ij})}_q = \sum_{q_i, q_j} (-1)^{k_i - k_j + q} (2k_{ij}+1)^{1/2} \begin{pmatrix} k_i & k_j & k_{ij} \\ q_i & q_j & -q \end{pmatrix} \underline{T}^{(k_i)}_{q_i}(\mathbf{s}_i) \underline{T}^{(k_j)}_{q_j}(\mathbf{s}_j) \quad . \tag{12}$$

Eq. (12) is also applicable to create higher-rank local operators (for $i = j$); local operators up to rank $k_i$ exist for $s_i = k_i / 2$. This work confines to $s_i = \frac{1}{2}$.

In a consecutive coupling scheme, $\underline{\mathbf{T}}^{(k_1)}(\mathbf{s}_1)$ and $\underline{\mathbf{T}}^{(k_2)}(\mathbf{s}_2)$ initially couple to $\underline{\mathbf{T}}^{k_1 k_2 (k_{12})}(\mathbf{s}_1, \mathbf{s}_2)$. Subsequently, $\underline{\mathbf{T}}^{k_1 k_2 (k_{12})}(\mathbf{s}_1, \mathbf{s}_2)$ is coupled with $\underline{\mathbf{T}}^{(k_3)}(\mathbf{s}_3)$ to $\underline{\mathbf{T}}^{k_1 k_2 k_{12} k_3 (k_{123})}(\mathbf{s}_1, \mathbf{s}_2, \mathbf{s}_3)$, and so on. The spin Hamiltonian $\underline{H}$ is isotropic, $[\underline{H}, \underline{S}_\alpha]$, $\alpha = x, y, z$, if it only encompasses terms of total rank $k = 0$. Conversely, $k > 0$ terms disrupt spin-rotational symmetry and thus introduce anisotropy. For the coefficients in an expansion in terms of spherical scalar products, Eq. (13),

$$\underline{H} = \sum_{k_1, k_2, \ldots, k_N} \sum_{k_{12}, k_{123}, \ldots, k} \sum_{q=-k}^{+k} (-1)^q X^{k_1 k_2 k_{12} \ldots k_N (k)}_{-q} \underline{T}^{k_1 k_2 k_{12} \ldots k_N (k)}_q \quad , \tag{13}$$

a simple trace formula can be deduced via successive decoupling involving Wigner-9$j$ symbols (note that $[X^{k_1 \ldots k_N (k)}_q]^* = (-1)^q X^{k_1 \ldots k_N (k)}_{-q}$),[40]

$$X^{k_1 k_2 k_{12} \ldots k_N (k)}_q = (-1)^k \times \mathrm{Tr}(\underline{T}^{k_1 k_2 k_{12} \ldots k_N (k)}_q \underline{H}) \times \prod_{i=1}^{N} \frac{(-1)^{k_i}(2k_i+1)}{\langle s_i \| \underline{\mathbf{T}}^{(k_i)} \| s_i \rangle^2} \quad . \tag{14}$$

Algebraic expressions for reduced matrix elements $\langle s_i \| \hat{\mathbf{T}}^{(k_i)} \| s_i \rangle$ are available in Ref. 68 (and elsewhere). In $s_i = \frac{1}{2}$ systems, the summation over local ITO ranks $k_i$ in Eq. (13) incorporates only $k_i = 0$ [$\underline{T}^0_0(\mathbf{s}_i) = 1$] or $k_i = 1$. To effectively characterize the $D = 2^N$ lowest levels of electronic Hamiltonians preserving spin symmetry, we only need to consider the $k=0$ terms in Eq. (14). Coefficients $X_{-q}^{k_1 k_2 k_{12} \ldots k_N (0)}$ for odd sums $K = \sum_i k_i$ vanish. In $s = \frac{1}{2}$ systems, $K$ equates to the number of interacting spin centers in the multi-exchange term $T_0^{k_1 k_2 k_{12} \ldots k_N (0)}$; for isotropic terms, $K$ is even $(K \leq N)$.[49] Counting the number $N_K^{(0)}$ of linearly independent operators $T_{\sim q}^{k_1 k_2 k_{12} \ldots k_N (0)}$ for a given $K$ is a combinatorial problem,

$$N_K^{(0)} = \binom{N}{K} R(K) , \qquad (15)$$

where $R(K)$ stands for the number of independent ways of coupling $K$ rank-1 operators to a total $k=0$. It could be helpful to envision $R(K)$ as the number of mutually orthogonal coupled $S=0$ states in a system of $K$ spin-1 sites. For systems with up to $N=10$ spin-1/2 centers, the numbers of independent terms for interactions among $K$ spins are compiled in Table 1.

Table 1: Numbers $N_K^{(0)}$ (Eq. (15)) of independent isotropic terms involving selections of $K$ sites within a system of $N$ spin-1/2 sites. The row sums in the final column represent the count of parameters of an isotropic spin Hamiltonian in the absence of PG symmetry.

| N | K | | | | | Total |
|---|---|---|---|---|---|---|
|   | 2 | 4 | 6 | 8 | 10 |   |
| 2 | 1 | 0 | 0 | 0 | 0 | 1 |
| 3 | 3 | 0 | 0 | 0 | 0 | 3 |
| 4 | 6 | 3 | 0 | 0 | 0 | 9 |
| 5 | 10 | 15 | 0 | 0 | 0 | 25 |
| 6 | 15 | 45 | 15 | 0 | 0 | 75 |
| 7 | 21 | 105 | 105 | 0 | 0 | 231 |
| 8 | 28 | 210 | 420 | 91 | 0 | 749 |
| 9 | 36 | 378 | 1260 | 819 | 0 | 2493 |
| 10 | 45 | 630 | 3150 | 4095 | 603 | 8523 |

In symmetric systems, the number of parameters is reduced due to the requirement that each independent term is totally symmetric with respect to the point group. Using $\Gamma_1$ as a generic label for the totally symmetric irreducible representation (irrep), the totally symmetric part of an operator is generated in Eq. (16).

$$T_{\sim\Gamma_1}^{k_1k_2k_{12}...k_N(0)} = \frac{1}{h}\sum_{g=1}^{h} G^\dagger(g) T_{\sim}^{k_1k_2k_{12}...k_N(0)} G(g) \quad (16)$$

Here $h$ denotes the order of the group, and $G(g)$ stands for a symmetry operator. Dimensions having specific PG labels and total spins (here corresponding to operator ranks $k$) have been reported for a few specific cases by Bärwinkel et al.,[50] who included local spin $s = 1$ (or local rank 1) in their analysis. These subspace dimensions can also be calculated with a code that simultaneously adapts the basis to $\hat{S}_z$ and PG symmetry.[51,52] We refrain from addressing the more complex question concerning the number of totally symmetric $k = 0$ operators with $K < N$ (for instance, 4c terms in the octahedron) in general terms, but specific cases are discussed in the following Results section.

## 3. Results

We initiate the analysis of our methodology for extracting the spin Hamiltonian from BS solutions with an examination of a symmetric dimer. Following this, we consider rings composed of three or four sites, and ultimately investigate an octahedron, see Figure 1.

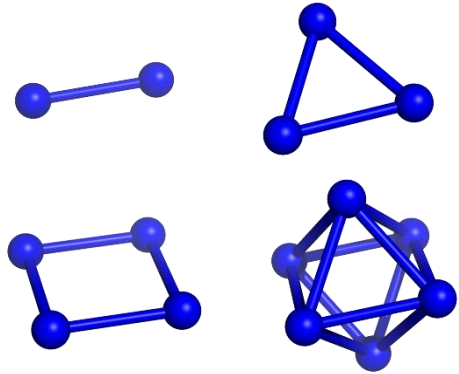

Figure 1: Illustration of the dimer, equilateral triangle, square, and octahedron, which are the systems explored in this study within the framework of the single-band Hubbard model. The aim is to compare $s = \frac{1}{2}$ Hamiltonians derived from the new BS method to their exact counterparts.

Spin-Hamiltonian parameters obtained through our new procedure are compared to results from canonical effective-Hamiltonian theory based on exact (FCI) solutions. We thereby illustrate that our approach can overcome inherent limitations of the prevalent mapping strategy that is solely dependent on mean-field energies. All numerical results in this paper are based on an on-site repulsion value set at $U = 1$.

**Dimer.** A single real parameter $\vartheta$ defines the $M = 0$ UHF solution $|\Phi\rangle$ for a dimer,[53]

$$|\Phi\rangle = \left[(\cos\vartheta)\hat{c}^\dagger_{g\uparrow} + (\sin\vartheta)\hat{c}^\dagger_{u\uparrow}\right]\left[(\cos\vartheta)\hat{c}^\dagger_{g\downarrow} - (\sin\vartheta)\hat{c}^\dagger_{u\downarrow}\right]|0\rangle, \quad (17)$$

where $\hat{c}^\dagger_{g\sigma} = \frac{1}{\sqrt{2}}\left(\hat{c}^\dagger_{1\sigma} + \hat{c}^\dagger_{2\sigma}\right)$ and $\hat{c}^\dagger_{u\sigma} = \frac{1}{\sqrt{2}}\left(\hat{c}^\dagger_{1\sigma} - \hat{c}^\dagger_{2\sigma}\right)$ are molecular orbitals that are even ($g$) or odd ($u$) under site exchange, respectively, and $\sigma = \uparrow, \downarrow$. The HF energy as a function of the variational variable $\vartheta$ and the parameters $U$ and $t$ is given in Eq. (18),

$$E_{HF} = \langle\Phi|\hat{H}|\Phi\rangle = \frac{U\cos^2(2\vartheta)}{2} - 2t\cos(2\vartheta). \quad (18)$$

For minimal-basis H$_2$, $E_{HF}$ depends on a larger number of parameters.[53] If $\frac{2t}{U} \geq 1$, Eq. (18) exhibits a single minimum $E_{RHF} = -2t + \frac{U}{2}$ at $\vartheta = 0$, representing a restricted HF (RHF) solution $|\Phi_{RHF}\rangle = \hat{c}^\dagger_{g\uparrow}\hat{c}^\dagger_{g\downarrow}|0\rangle$. A RHF $\to$ UHF instability, which is associated with the breaking of total-spin symmetry but conservation of the $z$-projection $M = 0$, appears when $\frac{2t}{U} < 1$. Consequently, Eq. (18) has two equivalent minima at $\vartheta = \pm\vartheta_0$, Eq. (19),

$$\vartheta_0 = \frac{1}{2}\tan^{-1}\frac{\sqrt{U^2 - 4t^2}}{2t}. \quad (19)$$

These minima correspond to UHF solutions $|\Uparrow\Downarrow\rangle$ and $|\Downarrow\Uparrow\rangle$,

$$E_{UHF} = -\frac{2t^2}{U}, \quad (20)$$

which break symmetry with respect to site permutation. For $M = +1$, HF results in the trivial exact solution $|\Uparrow\Uparrow\rangle = \hat{c}^\dagger_{1\uparrow}\hat{c}^\dagger_{2\uparrow}|0\rangle$ with total spin $S = 1$, $E_{HS} = 0$. In a unit system where $\hbar = 1$, the exact Heisenberg coupling constant $J_{ex} > 0$ (antiferromagnetic) aligns with the exact energy difference between the $S = 1$ level and the $S = 0$ ground state, Eq. (21).

$$J_{ex} = E(S = 1) - E(S = 0) \quad (21)$$

Eq. (22) provides an expansion of $J_{ex}$ in the strong-correlation regime (for small $\frac{t}{U}$).

$$J_{ex} = -\frac{U}{2} + \frac{1}{2}\sqrt{16t^2 + U^2} = \frac{4t^2}{U} - \frac{16t^4}{U^3} + \frac{128t^6}{U^5} + O[t^8/U^7] \quad (22)$$

The method known as Noodleman's approach (N) determines exchange couplings by fitting energies of different BS solutions. These solutions correspond to distinct spin configurations.

BS or HS states, such as $|\Uparrow\Downarrow\rangle$ or $|\Uparrow\Uparrow\rangle$, are associated with particular spin configurations, for instance, $|\uparrow\downarrow\rangle$ or $|\uparrow\uparrow\rangle$. For a dimer system, this association results in a linear system for the coupling constant $J_N$, as expressed in Eqs. (23) and (24),

$$E_{HS} = \frac{J_N}{4} + E_0, \qquad (23)$$

$$E_{BS} = -\frac{J_N}{4} + E_0, \qquad (24)$$

which yields

$$J_N = 2(E_{HS} - E_{BS}). \qquad (25)$$

The shift $E_0$ is employed to align the energy scales of the electronic model and the spin model. Notably, it is not a free parameter of the spin model; indeed, spin Hamiltonians are usually chosen to be traceless. The number of parameters extractable is clearly reduced by one relative to the count of symmetry-inequivalent spin configurations, and this latter number corresponds to the number of independent equations in a linear system, like that of Eqs. (23) and (24). For the Hubbard dimer, $J_N$, Eq. (26),

$$J_N = \frac{4t^2}{U}, \qquad (26)$$

is the leading contribution to the exact $J$, cf. Eq. (22).

As a refinement to Eq. (25), particularly in instances of strong antiferromagnetic coupling where the BS state displays significant spin pairing, it is advisable to employ the Yamaguchi formula, Eq. (27).[32,33]

$$J_Y = \frac{2(E_{HS} - E_{BS})}{\langle \hat{\mathbf{S}}^2 \rangle_{HS} - \langle \hat{\mathbf{S}}^2 \rangle_{BS}} \qquad (27)$$

$J_Y$ deviates from $J_N$ if $\langle \hat{\mathbf{S}}^2 \rangle_{HS}$ or $\langle \hat{\mathbf{S}}^2 \rangle_{BS}$ differ from their respective values in the corresponding spin configurations, that is, when $\langle \Uparrow\Uparrow | \hat{\mathbf{S}}^2 | \Uparrow\Uparrow \rangle \neq \langle \uparrow\uparrow | \hat{\mathbf{S}}^2 | \uparrow\uparrow \rangle$ or $\langle \Uparrow\Downarrow | \hat{\mathbf{S}}^2 | \Uparrow\Downarrow \rangle \neq \langle \uparrow\downarrow | \hat{\mathbf{S}}^2 | \uparrow\downarrow \rangle$. Specifically, $\langle \Uparrow\Downarrow | \hat{\mathbf{S}}^2 | \Uparrow\Downarrow \rangle$ is considerably smaller than $\langle \uparrow\downarrow | \hat{\mathbf{S}}^2 | \uparrow\downarrow \rangle$ if there is substantial spin pairing in $|\Uparrow\Downarrow\rangle$, where $|\Uparrow\Downarrow\rangle$ might be a UHF solution, or an unrestricted Kohn-Sham determinant in DFT, or a BS state calculated by coupled-cluster theory.[17] In the fully polarized configuration, deviations from the ideal value are typically much smaller, $\langle \Uparrow\Uparrow\Uparrow | \hat{S}^2 | \Uparrow\Uparrow\Uparrow \rangle \approx \langle \uparrow\uparrow | \hat{\underset{\sim}{S}}^2 | \uparrow\uparrow \rangle$. In the Hubbard dimer,

$$\langle \hat{S}^2 \rangle_{BS} = 1 - \frac{4t^2}{U^2} \; , \tag{28}$$

and $\langle \hat{S}^2 \rangle_{HS} = 2$ yield:

$$J_Y = \frac{4t^2/U}{1+(2t/U)^2} = \frac{4t^2}{U} - \frac{16t^4}{U^3} + \frac{64t^6}{U^5} + O[t^8/U^7] \; . \tag{29}$$

Eq. (29) demonstrates that $J_Y$ agrees with the $J_{ex}$ (Eq. (22)) through the first two expansion terms.

It can be easily shown that the new BS mapping strategy, introduced in the Theory section, delivers the effective Hamiltonian matrix $\underset{\sim}{\mathbf{H}}_{BS}$ in the basis $\{|\uparrow\uparrow\rangle, |\uparrow\downarrow\rangle, |\downarrow\uparrow\rangle, |\downarrow\downarrow\rangle\}$,

$$\underset{\sim}{\mathbf{H}}_{BS} = \begin{pmatrix} 0 & 0 & 0 & 0 \\ 0 & -\alpha & \alpha & 0 \\ 0 & \alpha & -\alpha & 0 \\ 0 & 0 & 0 & 0 \end{pmatrix}, \tag{30}$$

where $\alpha = 2t^2 U/(4t^2+U^2)$. By implementing a constant energy shift, a traceless spin Hamiltonian $\underset{\sim}{\mathbf{H}}_{BS}$ is obtained, Eq. (31),

$$\underset{\sim}{\mathbf{H}}_{BS} = \begin{pmatrix} \dfrac{J_{BS}}{4} & 0 & 0 & 0 \\ 0 & -\dfrac{J_{BS}}{4} & \dfrac{J_{BS}}{2} & 0 \\ 0 & \dfrac{J_{BS}}{2} & -\dfrac{J_{BS}}{4} & 0 \\ 0 & 0 & 0 & \dfrac{J_{BS}}{4} \end{pmatrix}, \tag{31}$$

where

$$J_{BS} = \frac{4t^2/U}{1+(2t/U)^2} = \frac{4t^2/U}{1+S} \; . \tag{32}$$

$\underset{\sim}{\mathbf{H}}_{BS}$ of Eq. (31) perfectly represents $\underset{\sim}{H}_{BS} = J_{BS} \underset{\sim}{\mathbf{s}}_1 \cdot \underset{\sim}{\mathbf{s}}_2$, that is, anisotropic artifacts that would split the $S=1$ level are absent. The overlap $S \equiv |\langle \underline{\Uparrow\Downarrow} | \underline{\Downarrow\Uparrow} \rangle|$ in Eq. (32) should not to be confused with the total-spin quantum number. In the Hubbard dimer, where $S = 1 - \langle \mathbf{S}^2 \rangle_{BS}$, our procedure aligns with the Yamaguchi correction, $J_{BS} = J_Y$, and with an alternative spin-contamination correction for BS states $|\Uparrow\Downarrow\rangle$ and $|\Downarrow\Uparrow\rangle$,[54]

$$J_F = \frac{2(E_{HS} - E_{BS})}{1+S}. \tag{33}$$

Unlike the Yamaguchi correction, which can be extended to multinuclear clusters,[33] it is not obvious how Eq. (33) could be generalized to systems with more than two spin centers, although our present work could be regarded to offer such a generalization.

In the molecular electronic structure problem, $\langle \hat{\mathbf{S}}^2 \rangle$ values do not directly relate to the overlap between BS states. Specifically, there is no straightforward relationship between the difference in spin expectation values $\langle \hat{\mathbf{S}}^2 \rangle_{HS} - \langle \hat{\mathbf{S}}^2 \rangle_{BS}$ and the overlap element $S \equiv |\langle \Uparrow\Downarrow | \Downarrow\Uparrow \rangle|$ (see, e.g., Ref. 54). Hence, predictions from either Eq. (27), Eq. (33), or our present procedure are not identical in realistic systems. However, the congruence with both previously proposed non-orthogonality/spin-contamination corrections of the Noodleman approach in the Hubbard dimer can be viewed as partial validation of the newly introduced mapping strategy.

**Triangle.** In a symmetric triangle, there exist two equivalent HS states, $|\Uparrow\Uparrow\Uparrow\rangle$ and $|\Downarrow\Downarrow\Downarrow\rangle$, and six equivalent BS states, related to $|\Uparrow\Uparrow\Downarrow\rangle$ by either cyclic site permutations (the full point-group is the dihedral group $D_3$) or by flipping all spins. The UHF solution $|\Uparrow\Uparrow\Downarrow\rangle$ in the Hubbard model (or a cluster of minimal-basis H atoms) depends on two real parameters,[55] and the analytical energy expression is relatively complicated (see Eq. 34 in Ref. 40), particularly compared to the dimer (Eq. (18)). Thus, our discussion focuses on numerical results. In UHF calculations, we self-consistently solve for the eigenfunctions of the single-particle Fock operator, $\hat{H}_{UHF}$, Eq. (34),

$$\hat{H}_{UHF} = U \sum_i \left( \hat{n}_{i\uparrow} \langle \hat{n}_{i\downarrow} \rangle + \langle \hat{n}_{i\uparrow} \rangle \hat{n}_{i\downarrow} - \langle \hat{n}_{i\uparrow} \rangle \langle \hat{n}_{i\downarrow} \rangle \right) - t \sum_{\langle i,j \rangle, \sigma} \hat{c}^\dagger_{i\sigma} \hat{c}_{j\sigma} , \tag{34}$$

which are occupied with $N$ electrons, $N_\uparrow + N_\downarrow = N$, that have a definite spin-projection on the $z$-axis. The fully covalent (non-ionic) spin configuration $|\Uparrow\Uparrow\Downarrow\rangle$ provides an appropriate initial guess to converge onto $|\Uparrow\Uparrow\Downarrow\rangle$. The Noodleman approach employs the energy difference between HS and BS states, Eq. (35),

$$J_N = E_{HS} - E_{BS}, \qquad (35)$$

while the generalized Yamaguchi correction[33] yields Eq. (36),[40]

$$J_Y = \frac{8(E_{HS} - E_{BS})}{15 - 4\langle \hat{\mathbf{S}}^2 \rangle_{BS}}. \qquad (36)$$

$J_{ex}$ can be expressed in terms of the difference between the excited term $^4A_1$ (totally symmetric in $D_3$, with total spin $S = \tfrac{3}{2}$) and the spatially degenerate ground term $^2E$ ($S = \tfrac{1}{2}$):

$$J_{ex} = \tfrac{2}{3}[E(S = \tfrac{3}{2}) - E(S = \tfrac{1}{2})]. \qquad (37)$$

$\utilde{H}_{BS}$ preserves $\utilde{\mathbf{S}}^2$ symmetry, but lacks full rotational invariance, leading to a splitting between the $M = \pm\tfrac{3}{2}$ and $M = \pm\tfrac{1}{2}$ levels of the $S = \tfrac{3}{2}$ multiplet. In Figure 2, we compare $J_{ex}$ to $J_N$ and $J_Y$ as a function of $t$ ($U = 1$); two different $J_{BS}$ values (a and b) are provided for our BS approach, which fit the energy difference (resulting from diagonalization of $\utilde{H}_{BS}$) between the $S = \tfrac{1}{2}$ ground state and either of the two artificially split excited levels, $|S = \tfrac{3}{2}, M = \pm\tfrac{3}{2}\rangle$ or $|S = \tfrac{3}{2}, M = \pm\tfrac{1}{2}\rangle$, respectively. Both $J_{BS(a)}$ and $J_{BS(b)}$ display better performance over the whole range compared to $J_Y$. On the other hand, $J_N$ does not follow the correct trend of decreasing $J/t^2$ ratios.

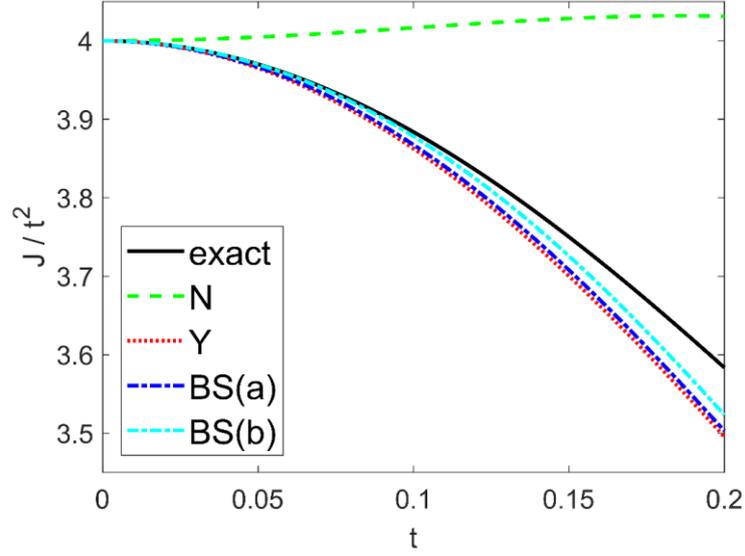

Figure 2: Coupling constants $J$ in the Hubbard triangle, from exact diagonalization, the N and Y approaches ($J_N$ and $J_Y$) and the present strategy (employing either the $\left|S=\tfrac{3}{2}, M=\pm\tfrac{3}{2}\right\rangle$ or $\left|S=\tfrac{3}{2}, M=\pm\tfrac{1}{2}\right\rangle$ levels to calculate $J_a$ and $J_b$, respectively, see main text) as a function of $t$ ($U=1$).

**Square.** As stated in Ref. 29, for four spin-1/2 sites, it is possible to construct three distinct isotropic four-center (4c) operators (see Table 1 in the Theory section). We define a shorthand notation $\utilde{T}^{k_{12}}$ for these operators:

$$\utilde{T}^{k_{12}} \equiv \left[\left[[\utilde{\mathbf{T}}^{(1)}(\mathbf{s}_1) \otimes \utilde{\mathbf{T}}^{(1)}(\mathbf{s}_2)]^{(k_{12})} \otimes \utilde{\mathbf{T}}^{(1)}(\mathbf{s}_3)\right]^{(1)} \otimes \utilde{\mathbf{T}}^{(1)}(\mathbf{s}_4)\right]_0^{(0)} . \tag{38}$$

By employing Eqs. (39) and (40),

$$\left[\utilde{\mathbf{T}}^{(1)}(\mathbf{s}_i) \otimes \utilde{\mathbf{T}}^{(1)}(\mathbf{s}_j)\right]_0^{(0)} = -\frac{1}{\sqrt{3}} \utilde{\mathbf{s}}_i \cdot \utilde{\mathbf{s}}_j , \tag{39}$$

$$\left[\utilde{\mathbf{T}}^{(1)}(\mathbf{s}_1) \otimes \utilde{\mathbf{T}}^{(1)}(\mathbf{s}_2)\right]_q^{(1)} = \frac{i}{\sqrt{2}} (\utilde{\mathbf{s}}_1 \times \utilde{\mathbf{s}}_2)_q , \tag{40}$$

we can recast $\utilde{T}^{k_{12}=0}$ and $\utilde{T}^{k_{12}=1}$ as Eqs. (41) and (42),

$$\utilde{T}^{k_{12}=0} = -\frac{1}{3}(\utilde{\mathbf{s}}_1 \cdot \utilde{\mathbf{s}}_2)(\utilde{\mathbf{s}}_3 \cdot \utilde{\mathbf{s}}_4) , \tag{41}$$

$$\utilde{T}^{k_{12}=1} = \frac{1}{2\sqrt{3}}\left[(\utilde{\mathbf{s}}_1 \times \utilde{\mathbf{s}}_2) \times \utilde{\mathbf{s}}_3\right] \cdot \utilde{\mathbf{s}}_4 = \frac{1}{2\sqrt{3}}\left[(\utilde{\mathbf{s}}_1 \cdot \utilde{\mathbf{s}}_3)(\utilde{\mathbf{s}}_2 \cdot \utilde{\mathbf{s}}_4) - (\utilde{\mathbf{s}}_1 \cdot \utilde{\mathbf{s}}_4)(\utilde{\mathbf{s}}_2 \cdot \utilde{\mathbf{s}}_3)\right] . \tag{42}$$

However, neither operator $\undertilde{T}^{k_{12}}$ is symmetric with respect to the $D_4$ point group of the square (sites are numbered consecutively). Indeed, there exist only two linearly independent, isotropic, and totally symmetric 4c operators in the $s = \frac{1}{2}$ square (see below).[56]

Through fourth order in quasi-degenerate perturbation theory (PT4[29]), the spin Hamiltonian of Eq. (43) is obtained,

$$H_{PT4} = J_{12}(\undertilde{s}_1 \cdot \undertilde{s}_2 + \undertilde{s}_2 \cdot \undertilde{s}_3 + \undertilde{s}_3 \cdot \undertilde{s}_4 + \undertilde{s}_4 \cdot \undertilde{s}_1)$$
$$+ J_{13}(\undertilde{s}_1 \cdot \undertilde{s}_3 + \undertilde{s}_2 \cdot \undertilde{s}_4) + J_{cyc}\undertilde{Q}_{cyc} \quad , \tag{43}$$

with parameters

$$J_{12} = \frac{4t^2}{U} - \frac{20t^4}{U^3} \quad , \tag{44}$$

$$J_{13} = \frac{4t^4}{U^3} \quad , \tag{45}$$

and

$$J_{cyc} = \frac{80t^4}{U^3} \quad . \tag{46}$$

The 4c operator $\undertilde{Q}_{cyc}$, Eq. (47),

$$\undertilde{Q}_{cyc} = (\undertilde{s}_1 \cdot \undertilde{s}_2)(\undertilde{s}_3 \cdot \undertilde{s}_4) + (\undertilde{s}_1 \cdot \undertilde{s}_4)(\undertilde{s}_2 \cdot \undertilde{s}_3) - (\undertilde{s}_1 \cdot \undertilde{s}_3)(\undertilde{s}_2 \cdot \undertilde{s}_4) \quad , \tag{47}$$

is called cyclic exchange, because it emerges from the symmetrized cyclical exchange $\undertilde{P}_{1234}$ of the four spins,[57] see, e.g., Eq. 4 in Ref. 58,

$$\undertilde{P}_{1234} + \undertilde{P}_{1234}^{-1} = 4(\undertilde{s}_1 \cdot \undertilde{s}_2)(\undertilde{s}_3 \cdot \undertilde{s}_4) + 4(\undertilde{s}_1 \cdot \undertilde{s}_4)(\undertilde{s}_2 \cdot \undertilde{s}_3) - 4(\undertilde{s}_1 \cdot \undertilde{s}_3)(\undertilde{s}_2 \cdot \undertilde{s}_4) + \sum_{i<j} \undertilde{s}_i \cdot \undertilde{s}_j + \frac{1}{4} \quad , \tag{48}$$

upon neglecting the pairwise terms and the constant. Note that $\undertilde{Q}_{cyc}$ represents a linear combination of $\undertilde{T}^{k_{12}=0}$ (Eq. (41)) and $\undertilde{T}^{k_{12}=1}$ (Eq. (42)):

$$\undertilde{Q}_{cyc} = -3\undertilde{T}^{k_{12}=0} - 2\sqrt{3}\undertilde{T}^{k_{12}=1} \quad . \tag{49}$$

However, another 4c term exists that is compatible with $D_4$ symmetry:

$$\undertilde{Q}_{noncyc} = (\undertilde{s}_1 \cdot \undertilde{s}_2)(\undertilde{s}_3 \cdot \undertilde{s}_4) + (\undertilde{s}_1 \cdot \undertilde{s}_4)(\undertilde{s}_2 \cdot \undertilde{s}_3) + 6(\undertilde{s}_1 \cdot \undertilde{s}_3)(\undertilde{s}_2 \cdot \undertilde{s}_4) \quad . \tag{50}$$

$\undertilde{Q}_{cyc}$ and $\undertilde{Q}_{noncyc}$ are independent, signified by the vanishing trace of their product, Eq. (51):

$$\text{Tr}(\undertilde{Q}_{cyc}\undertilde{Q}_{noncyc}) = 0 \quad . \tag{51}$$

Our (non-exhaustive) exploration of the literature did not provide any indication that the "non-cyclic" exchange has previously been considered in any theoretical or experimental study. Overall, the complete spin Hamiltonian for the square in Eq. (52) contains four independent parameters,

$$\underset{\sim}{H} = J_{12}(\underset{\sim}{\mathbf{s}}_1 \cdot \underset{\sim}{\mathbf{s}}_2 + \underset{\sim}{\mathbf{s}}_2 \cdot \underset{\sim}{\mathbf{s}}_3 + \underset{\sim}{\mathbf{s}}_3 \cdot \underset{\sim}{\mathbf{s}}_4 + \underset{\sim}{\mathbf{s}}_4 \cdot \underset{\sim}{\mathbf{s}}_1) \\ + J_{13}(\underset{\sim}{\mathbf{s}}_1 \cdot \underset{\sim}{\mathbf{s}}_3 + \underset{\sim}{\mathbf{s}}_2 \cdot \underset{\sim}{\mathbf{s}}_4) + J_{cyc}\underset{\sim}{O}_{cyc} + J_{noncyc}\underset{\sim}{O}_{noncyc} \quad . \tag{52}$$

The following trace formulas allow to extract $J_{12}$, $J_{13}$, $J_{cyc}$ and $J_{noncyc}$ from an effective Hamiltonian $\underset{\sim}{H}$ that maintains $D_4$ symmetry:

$$J_{12} = \frac{1}{3}\text{Tr}\left[(\underset{\sim}{\mathbf{s}}_1 \cdot \underset{\sim}{\mathbf{s}}_2)\underset{\sim}{H}\right] \tag{53}$$

$$J_{13} = \frac{1}{3}\text{Tr}\left[(\underset{\sim}{\mathbf{s}}_1 \cdot \underset{\sim}{\mathbf{s}}_3)\underset{\sim}{H}\right] \tag{54}$$

$$J_{cyc} = \frac{16}{21}\text{Tr}\left[\underset{\sim}{O}_{cyc}\underset{\sim}{H}\right] \tag{55}$$

$$J_{noncyc} = \frac{4}{105}\text{Tr}\left[\underset{\sim}{O}_{noncyc}\underset{\sim}{H}\right] \quad . \tag{56}$$

In Figure 5, we present a comparison of spin-multiplet spectra obtained from exact diagonalization and our BS methodology. The alignment is rather close.

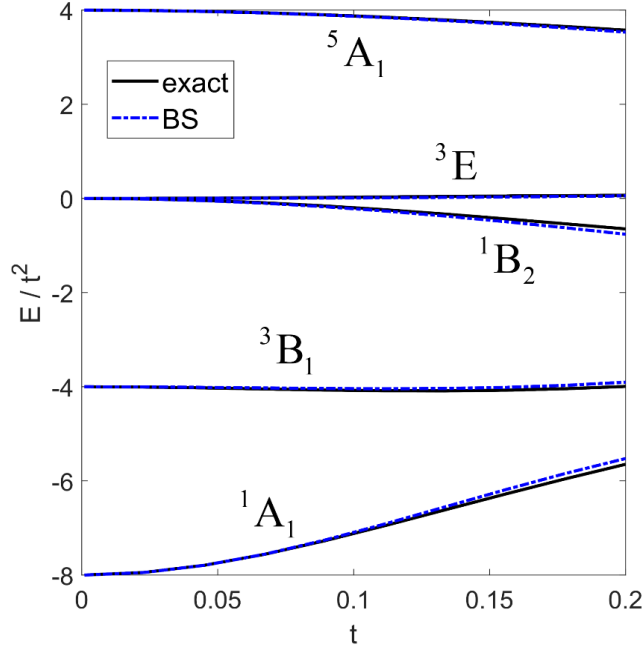

Figure 3: Spin-multiplet spectra derived from exact diagonalization (including the 16 lowest eigenstates of the half-filled single-band Hubbard square; levels were shifted to have a mean energy of zero) compared to our BS approach, as a function of $t$ ($U = 1$). Levels are labeled by term symbols which denote the spin multiplicity (given by the left superscript) and identify the Mulliken label for the $D_4$ point group.

If one follows the Noodleman approach, the energies of the four inequivalent mean-field solutions corresponding to spin configurations $|\uparrow\uparrow\uparrow\uparrow\rangle$, $|\uparrow\uparrow\uparrow\downarrow\rangle$, $|\uparrow\uparrow\downarrow\downarrow\rangle$ and $|\uparrow\downarrow\uparrow\downarrow\rangle$ set up an underdetermined system for the four model parameters. As performed in Ref. 29, one may decide to ignore $J_{\text{noncyc}}$, based on $\underline{H}_{\text{PT4}}$ (Eq. (43)). This provides the linear system of Eq. (57).

$$\begin{aligned}
E_{\uparrow\uparrow\uparrow\uparrow} &= J_{12} + \tfrac{1}{2} J_{13} + \tfrac{1}{16} J_{\text{cyc}} + E_0 \\
E_{\uparrow\uparrow\uparrow\downarrow} &= -\tfrac{1}{16} J_{\text{cyc}} + E_0 \\
E_{\uparrow\uparrow\downarrow\downarrow} &= -\tfrac{1}{2} J_{13} + \tfrac{1}{16} J_{\text{cyc}} + E_0 \\
E_{\uparrow\downarrow\uparrow\downarrow} &= -J_{12} + \tfrac{1}{2} J_{13} + \tfrac{1}{16} J_{\text{cyc}} + E_0
\end{aligned} \quad (57)$$

It is worth noting that, as the diagonal elements of $\underline{T}^{k_{12}=1}$ vanish in the uncoupled spin basis, the diagonal elements of $\underline{Q}_{\text{cyc}}$, which generate the $J_{\text{cyc}}$ terms in Eq. (57), just count the $(-3\underline{T}^{k_{12}=0})$ contribution, cf. Eq. (49). The solution to Eq. (57) is given in Eq. (58).

$$\begin{aligned}
J_{12} &= \tfrac{1}{2}(E_{\uparrow\uparrow\uparrow\uparrow} - E_{\uparrow\downarrow\uparrow\downarrow}) \\
J_{13} &= \tfrac{1}{2}(E_{\uparrow\uparrow\uparrow\uparrow} + E_{\uparrow\downarrow\uparrow\downarrow}) - E_{\uparrow\uparrow\downarrow\downarrow} \\
J_{cyc} &= 2(E_{\uparrow\uparrow\uparrow\uparrow} + E_{\uparrow\downarrow\uparrow\downarrow}) - 8 E_{\uparrow\uparrow\uparrow\downarrow} + 4 E_{\uparrow\uparrow\downarrow\downarrow} \\
E_0 &= \tfrac{1}{8}(E_{\uparrow\uparrow\uparrow\uparrow} + E_{\uparrow\downarrow\uparrow\downarrow}) + \tfrac{1}{2} E_{\uparrow\uparrow\uparrow\downarrow} + \tfrac{1}{4} E_{\uparrow\uparrow\downarrow\downarrow}
\end{aligned} \qquad (58)$$

To the best of our knowledge, the generalized Yamaguchi correction was never deployed to quantify four-spin interactions and we make no attempt to integrate them into this scheme. In any case, the Yamaguchi correction still would not allow for the calculation of all four parameters. Our primary intention here is to illustrate that our new BS approach allows us to obtain all parameters in an unbiased manner.

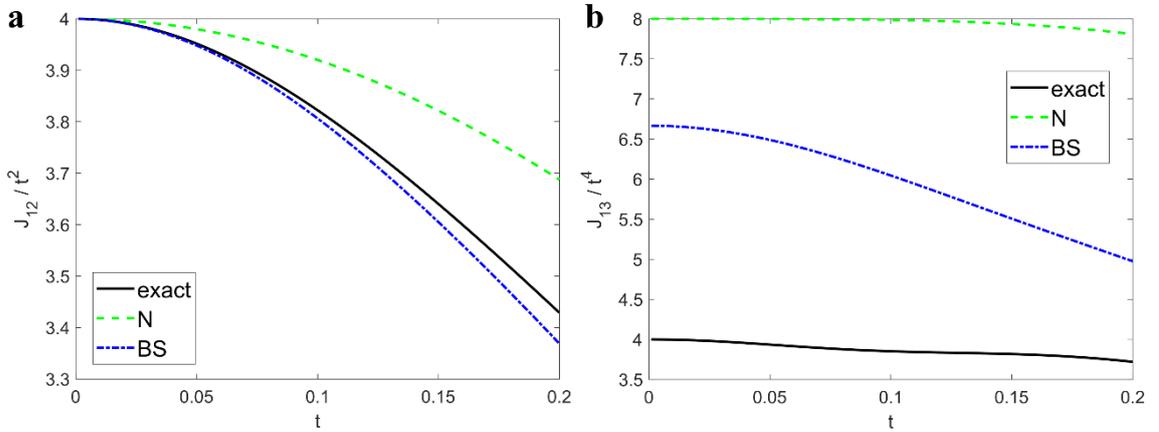

Figure 4: Coupling constants $J_{12}$ (**a**) and $J_{13}$ (**b**) in the square, from exact diagonalization, the N approach and the present method, as a function of $t$ ($U = 1$).

The nearest-neighbor (NN) and next-nearest neighbor (NNN) couplings $J_{12}$ and $J_{13}$, respectively, as determined from exact diagonalization, are compared with the N approach and our novel BS strategy in Figure 4. For $t \to 0$, N and BS correctly converge to $J_{12} = 4t^2$, while for $J_{13}$ both deviate in this limit; notably, the N approach overestimates $J_{13}$ by a factor of 2. The BS strategy offers a refinement for $J_{13}$, and while there is a relatively large error, it has minimal impact on the energy spectra. For $J_{cyc}$ (Figure 5), the N approach is apparently exact as $t$ approaches zero, see Figure 5. In this limit, while our BS approach correctly reproduces the fourth-order scaling, it does not yield the accurate $J/t^4$ ratio. However, the N approach results in an incorrect (positive) slope, whereas our method aligns more closely with the exact result in the intermediate-coupling regime, where cyclic exchange becomes crucial for describing energy levels.

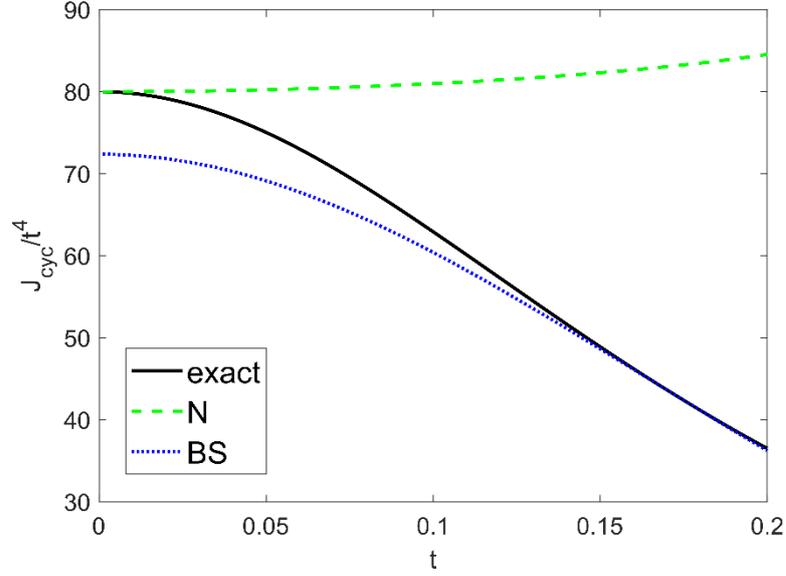

Figure 5: Cyclic exchange $J_{cyc}$ in the square, from exact diagonalization, the N approach and the present method, as a function of $t$ ($U = 1$).

In our exploration of the non-cyclic exchange, we found that the BS approach demonstrates a fourth-power scaling with $t$ (Figure 6b). In contrast, exact diagonalization indicates a sixth-order scaling (Figure 6a). However, as we approach the intermediate-coupling regime, the alignment in terms of numerical values becomes more reasonable. For instance, at $t = 0.2$, the BS approach predicts $J_{noncyc} = -2.03 \times 10^{-3}$, close to the exact value of $J_{noncyc} = -1.60 \times 10^{-3}$.

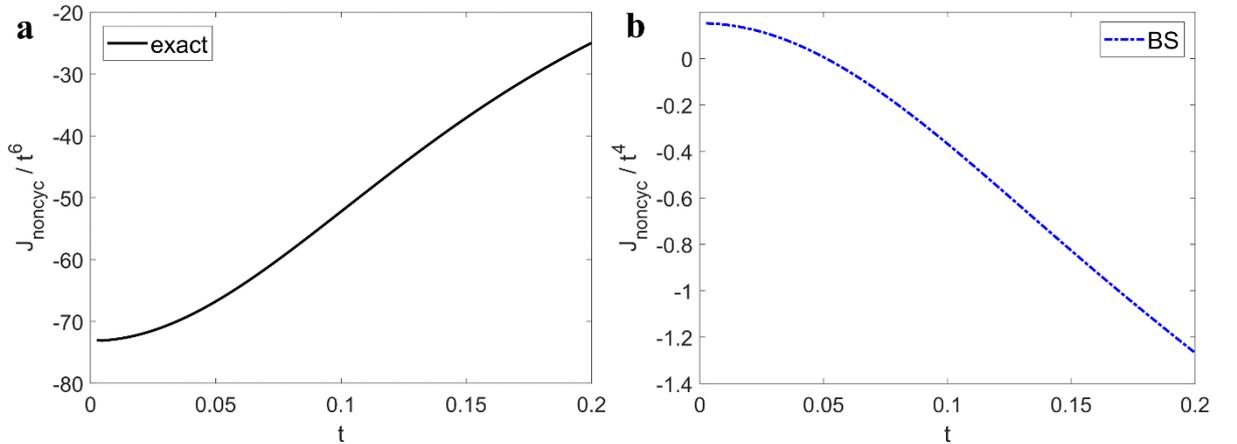

Figure 6: Non-cyclic exchange $J_{cyc}$ in the square, from exact diagonalization (**a**), and the present method (**b**), as a function of $t$ ($U = 1$). Note that the exact and BS values have leading sixth- and fourth-order contributions, respectively, see *y*-axis labels and main text.

The discrepancy could be perceived as less stark if we opted for a different linear combination of the 4c operators (cf. our choice of 4c operators in the octahedron, Eqs. (64), (65) and (66) below), because such a selection results in both terms showing a fourth-order scaling. In real magnetic compounds, it is not guaranteed that the cyclic exchange will vastly outweigh non-cyclic exchange, and having an unbiased methodology that can probe both is advantageous. As an example, non-cyclic exchange emerges at PT4, rather than PT6, when NNN hopping is introduced. With an NNN hopping parameter $t' = \frac{t}{2}$, our PT4 calculation (using symbolic algebra in `Matlab`) yields:

$$J_{12} = \frac{4t^2}{U} - \frac{20t^4}{U^3}, \tag{59}$$

$$J_{13} = \frac{t^2}{U} + \frac{t^4}{U^3}, \tag{60}$$

$$J_{cyc} = \frac{520}{7} \frac{t^4}{U^3}, \tag{61}$$

$$J_{noncyc} = \frac{40}{7} \frac{t^4}{U^3}. \tag{62}$$

The exact values of $J_{cyc}$ and $J_{noncyc}$ are compared to those determined through our BS strategy in Figure 7, showcasing consistent agreement across the entire range.

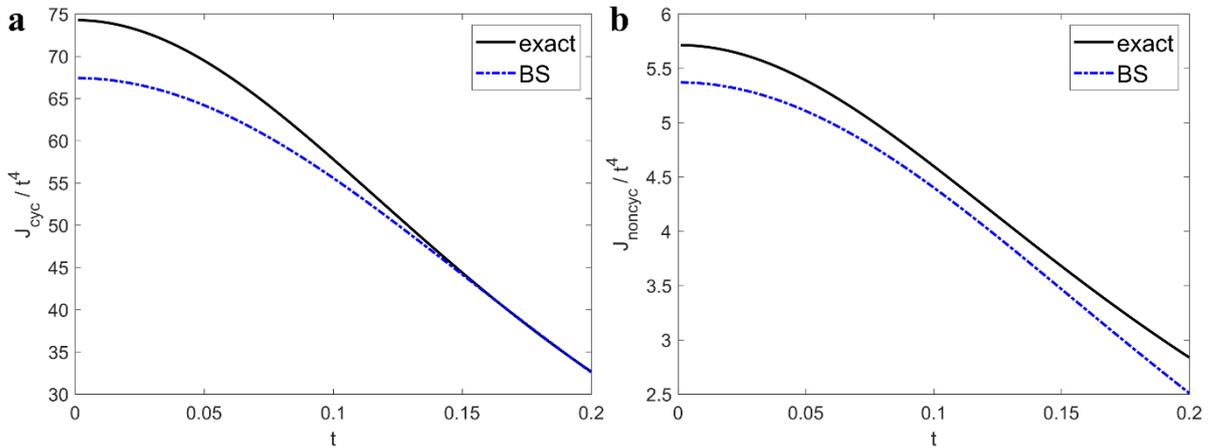

Figure 7: Four-center exchange $J_{cyc}$ (**a**) and $J_{noncyc}$ (**b**) in a square with NNN hopping $t' = \frac{t}{2}$, from exact diagonalization and the present method, as a function of $t$ ($U = 1$).

Unlike the dimer and the triangle, $\underset{\sim}{H}_{BS}$ breaks $\underset{\sim}{S}^2$ symmetry. However, adapting to $\underset{\sim}{S}^2$ (as given by Eq. (9)) has a negligible effect on the extracted parameter values. We checked that the eigenvalues of $\underset{\sim}{H}_{BS}$ (without the $\underset{\sim}{S}^2$ adaptation) align closely (within the precision of numerical double accuracy) with the eigenvalues of the generalized eigenvalue problem in the space of the mean-field solutions, Eq. (63),

$$\mathbf{HC} = \mathbf{SCE} \ , \tag{63}$$

where $H_{ij} \equiv \langle \Phi_i | \hat{H} | \Phi_j \rangle$, $S_{ij} \equiv \langle \Phi_i | \Phi_j \rangle$, $\mathbf{E}$ is a diagonal energy matrix, and the columns of $\mathbf{C}$ consist of the generalized eigenvectors. Put differently, $\underset{\sim}{H}_{BS}$ replicates the energy spectrum of the NOCI problem. From this perspective, one might consider building an effective Hamiltonian using the canonical procedure (see Theory section) based on NOCI eigenvectors. However, if only a limited set of BS states is accessible, constructing the effective Hamiltonian directly from the BS solutions via a reduced scheme might be conceptually simpler than tackling the NOCI problem in an incomplete basis. The development of such a reduced scheme, which uses only a subset of the BS states' energies and off-diagonal matrix elements is outside the scope of this work.

**Octahedron.** The $O_h$ symmetry reduces the count of two-, four-, and six-center terms from 15, 45, and 15 (as detailed in Table 1) to 2, 4, and 3, respectively. This results in a total of 9 independent parameters. Rather than employing the cyclic or non-cyclic exchange terminology, we construct the 4c terms in a manner that might be perceived as more intuitive, see Figure 8. The site numbering is defined in Figure 8a.

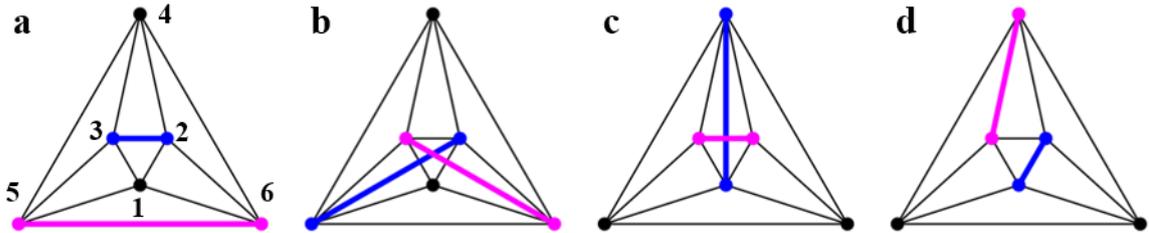

Figure 8: Illustration of the 4c-term construction for the $s = \tfrac{1}{2}$ octahedron. The six vertices are projected onto the plane, and black connecting lines correspond to edges. Plot **a** establishes the numbering of the six centers. Each scalar coupling $\underset{\sim}{\mathbf{s}}_i \cdot \underset{\sim}{\mathbf{s}}_j$ is indicated by a bold line (blue or pink), overall yielding a 4c operator $(\underset{\sim}{\mathbf{s}}_i \cdot \underset{\sim}{\mathbf{s}}_j)(\underset{\sim}{\mathbf{s}}_k \cdot \underset{\sim}{\mathbf{s}}_l)$. From $O_h$ symmetrization, the four operators $\underset{\sim}{A}_4$ (Eq. (64)), $\underset{\sim}{A}'_4$ (Eq. (65)), $\underset{\sim}{C}_4$ (Eq. (67)), and $\underset{\sim}{C}'_4$ (Eq. (68)) are obtained.

We first exclude a distant pair (NNN) and form two edge pairs (NN) from the remaining four sites (Figure 8a),

$$A_4 = \left[ (\underline{s}_2 \cdot \underline{s}_3)(\underline{s}_5 \cdot \underline{s}_6) \right]_{A_{1g}} =$$
$$\tfrac{1}{6} \left[ (\underline{s}_2 \cdot \underline{s}_3)(\underline{s}_5 \cdot \underline{s}_6) + (\underline{s}_3 \cdot \underline{s}_5)(\underline{s}_2 \cdot \underline{s}_6) \right.$$
$$+ (\underline{s}_1 \cdot \underline{s}_2)(\underline{s}_4 \cdot \underline{s}_5) + (\underline{s}_1 \cdot \underline{s}_5)(\underline{s}_2 \cdot \underline{s}_4)$$
$$\left. + (\underline{s}_1 \cdot \underline{s}_3)(\underline{s}_4 \cdot \underline{s}_6) + (\underline{s}_1 \cdot \underline{s}_6)(\underline{s}_3 \cdot \underline{s}_4) \right]$$
(64)

In Eq. (64), $A_{1g}$ denotes the totally symmetric irrep of $O_h$ (corresponding to the generic $\Gamma_1$ label in Eq. (16)). Forming NNN pairs in a topologically identical selection of four centers yield another term (Figure 8b),

$$A'_4 = \left[ (\underline{s}_2 \cdot \underline{s}_5)(\underline{s}_3 \cdot \underline{s}_6) \right]_{A_{1g}} =$$
$$\tfrac{1}{3} \left[ (\underline{s}_2 \cdot \underline{s}_5)(\underline{s}_3 \cdot \underline{s}_6) + (\underline{s}_1 \cdot \underline{s}_4)(\underline{s}_2 \cdot \underline{s}_5) + (\underline{s}_1 \cdot \underline{s}_4)(\underline{s}_3 \cdot \underline{s}_6) \right]$$
(65)

$A_4$ and $A'_4$ are not independent, that is, $\mathrm{Tr}(A_4 A'_4) \neq 0$. Thus, $A'_4$ should be "orthogonalized", as in Eq. (66),

$$B_4 = A'_4 - \frac{\mathrm{Tr}(A'_4 A_4)}{\mathrm{Tr}(A_4 A_4)} A_4 = A'_4 - \tfrac{1}{2} A_4 .$$
(66)

Another type of 4c interaction omits an edge pair (NN) while forming either one edge and one distant pair or two edge pairs. $C_4$ in Eq. (67) corresponds to the first option (Figure 8c):

$$C_4 = \left[ (\underline{s}_1 \cdot \underline{s}_4)(\underline{s}_2 \cdot \underline{s}_3) \right]_{A_{1g}} .$$
(67)

$C'_4$ in Eq. (68) corresponds to the second option (Figure 8d):

$$C'_4 = \left[ (\underline{s}_1 \cdot \underline{s}_2)(\underline{s}_3 \cdot \underline{s}_4) \right]_{A_{1g}} .$$
(68)

For brevity, the expansions of the symmetrized operators $C_4$ and $C'$ are not reported here. Orthogonalization yields:

$$D_4 = C'_4 - \tfrac{1}{3} C .$$
(69)

$C_4$ and $D_4$ are orthogonal to $A_4$ and $B_4$, so $\{A_4, B_4, C_4, D_4\}$ form a complete set of independent totally symmetric isotropic 4c-operators in the $s = \tfrac{1}{2}$ octahedron.

The construction of 6c terms is illustrated in Figure 9. A product of distant-site scalar couplings (Figure 9a) yields the totally symmetric $A_6$, Eq. (70):

$$A_6 = \left[ (\underline{s}_1 \cdot \underline{s}_4)(\underline{s}_2 \cdot \underline{s}_5)(\underline{s}_3 \cdot \underline{s}_6) \right]_{A_{1g}} = (\underline{s}_1 \cdot \underline{s}_4)(\underline{s}_2 \cdot \underline{s}_5)(\underline{s}_3 \cdot \underline{s}_6) .$$
(70)

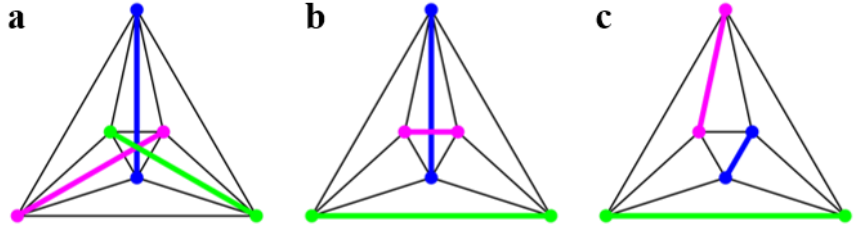

Figure 9: Illustration of the 6c-term construction for an $s=\frac{1}{2}$ octahedron. Each scalar coupling $\underline{s}_i \cdot \underline{s}_j$ is indicated by a bold colored line, overall yielding a 6c operator $(\underline{s}_i \cdot \underline{s}_j)(\underline{s}_k \cdot \underline{s}_l)(\underline{s}_m \cdot \underline{s}_n)$. While the first operator (**a**) is already totally symmetric, symmetrization of the other two (**b** and **c**) yields $\underline{A}'_6$ (Eq. (71)) and $\underline{A}''_6$ (Eq. (73)).

Alternatively, with one distant and two edge pairs (Figure 9b):

$$\begin{aligned}
\underline{A}'_6 &= \left[ (\underline{s}_1 \cdot \underline{s}_4)(\underline{s}_2 \cdot \underline{s}_3)(\underline{s}_5 \cdot \underline{s}_6) \right]_{A_{1g}} = \\
&\tfrac{1}{6} \big[ (\underline{s}_1 \cdot \underline{s}_4)(\underline{s}_2 \cdot \underline{s}_3)(\underline{s}_5 \cdot \underline{s}_6) + (\underline{s}_1 \cdot \underline{s}_4)(\underline{s}_2 \cdot \underline{s}_6)(\underline{s}_3 \cdot \underline{s}_5) \\
&+ (\underline{s}_2 \cdot \underline{s}_5)(\underline{s}_1 \cdot \underline{s}_3)(\underline{s}_4 \cdot \underline{s}_6) + (\underline{s}_2 \cdot \underline{s}_5)(\underline{s}_1 \cdot \underline{s}_6)(\underline{s}_3 \cdot \underline{s}_4) \\
&+ (\underline{s}_3 \cdot \underline{s}_6)(\underline{s}_1 \cdot \underline{s}_2)(\underline{s}_4 \cdot \underline{s}_5) + (\underline{s}_3 \cdot \underline{s}_6)(\underline{s}_1 \cdot \underline{s}_5)(\underline{s}_2 \cdot \underline{s}_4) \big]
\end{aligned} \quad (71)$$

$$B_6 = \underline{A}'_6 - \frac{\mathrm{Tr}(\underline{A}'_6 \underline{A}_6)}{\mathrm{Tr}(\underline{A}_6 \underline{A}_6)} \underline{A}_6 = \underline{A}'_6 - \tfrac{1}{3} \underline{A}_6 . \quad (72)$$

Lastly, with three edge pairs (Figure 9c),

$$\begin{aligned}
\underline{A}''_6 &= \left[ (\underline{s}_1 \cdot \underline{s}_2)(\underline{s}_3 \cdot \underline{s}_4)(\underline{s}_5 \cdot \underline{s}_6) \right]_{A_{1g}} \\
&\tfrac{1}{8} \big\{ (\underline{s}_1 \cdot \underline{s}_2) \left[ (\underline{s}_3 \cdot \underline{s}_4)(\underline{s}_5 \cdot \underline{s}_6) + (\underline{s}_3 \cdot \underline{s}_5)(\underline{s}_4 \cdot \underline{s}_6) \right] \\
&+ (\underline{s}_1 \cdot \underline{s}_3) \left[ (\underline{s}_2 \cdot \underline{s}_4)(\underline{s}_5 \cdot \underline{s}_6) + (\underline{s}_2 \cdot \underline{s}_6)(\underline{s}_4 \cdot \underline{s}_5) \right] \\
&+ (\underline{s}_1 \cdot \underline{s}_5) \left[ (\underline{s}_2 \cdot \underline{s}_6)(\underline{s}_3 \cdot \underline{s}_4) + (\underline{s}_2 \cdot \underline{s}_3)(\underline{s}_4 \cdot \underline{s}_6) \right] \\
&+ (\underline{s}_1 \cdot \underline{s}_6) \left[ (\underline{s}_2 \cdot \underline{s}_4)(\underline{s}_3 \cdot \underline{s}_5) + (\underline{s}_2 \cdot \underline{s}_3)(\underline{s}_4 \cdot \underline{s}_5) \right] \big\}
\end{aligned} \quad (73)$$

which, upon orthogonalization with respect to $\underline{A}_6$ and $\underline{B}_6$ yields:

$$C_6 = \underline{A}''_6 - \frac{\mathrm{Tr}(\underline{A}''_6 \underline{A}_6)}{\mathrm{Tr}(\underline{A}_6 \underline{A}_6)} \underline{A}_6 - \frac{\mathrm{Tr}(\underline{A}''_6 \underline{B}_6)}{\mathrm{Tr}(\underline{B}_6 \underline{B}_6)} \underline{B}_6 = \underline{A}''_6 - \tfrac{1}{11} \underline{A}_6 - \underline{B}_6 . \quad (74)$$

The isotropic Hamiltonian of Eq. (75) results for the $s=\frac{1}{2}$ octahedron.

$$\begin{aligned}
\underset{\sim}{H} = J_{12}(&\underset{\sim}{S}_1 \cdot \underset{\sim}{S}_2 + \underset{\sim}{S}_1 \cdot \underset{\sim}{S}_3 + \underset{\sim}{S}_1 \cdot \underset{\sim}{S}_5 + \underset{\sim}{S}_1 \cdot \underset{\sim}{S}_6 + \\
&\underset{\sim}{S}_2 \cdot \underset{\sim}{S}_3 + \underset{\sim}{S}_2 \cdot \underset{\sim}{S}_4 + \underset{\sim}{S}_2 \cdot \underset{\sim}{S}_6 + \underset{\sim}{S}_3 \cdot \underset{\sim}{S}_4 + \\
&\underset{\sim}{S}_3 \cdot \underset{\sim}{S}_5 + \underset{\sim}{S}_4 \cdot \underset{\sim}{S}_5 + \underset{\sim}{S}_4 \cdot \underset{\sim}{S}_6 + \underset{\sim}{S}_5 \cdot \underset{\sim}{S}_6) \\
+ J_{14}(&\underset{\sim}{S}_1 \cdot \underset{\sim}{S}_4 + \underset{\sim}{S}_2 \cdot \underset{\sim}{S}_5 + \underset{\sim}{S}_3 \cdot \underset{\sim}{S}_6) \\
+ a_4 &\underset{\sim}{A}_4 + b_4 \underset{\sim}{B}_4 + c_4 \underset{\sim}{C}_4 + d_4 \underset{\sim}{D}_4 \\
+ a_6 &\underset{\sim}{A}_6 + b_6 \underset{\sim}{B}_6 + c_6 \underset{\sim}{C}_6
\end{aligned} \qquad (75)$$

The following trace equations allow to extract the parameters if $\underset{\sim}{H}$ preserves $O_h$ symmetry:

$$J_{12} = \tfrac{1}{12}\mathrm{Tr}\left[\underset{\sim}{H}(\underset{\sim}{S}_1 \cdot \underset{\sim}{S}_2)\right] \qquad (76)$$

$$J_{14} = \tfrac{1}{12}\mathrm{Tr}\left[\underset{\sim}{H}(\underset{\sim}{S}_1 \cdot \underset{\sim}{S}_4)\right] \qquad (77)$$

$$a_4 = 2\mathrm{Tr}(\underset{\sim}{H}\underset{\sim}{A}_4) \qquad (78)$$

$$b_4 = \tfrac{8}{5}\mathrm{Tr}(\underset{\sim}{H}\underset{\sim}{B}_4) \qquad (79)$$

$$c_4 = \tfrac{16}{3}\mathrm{Tr}(\underset{\sim}{H}\underset{\sim}{C}_4) \qquad (80)$$

$$d_4 = \tfrac{48}{5}\mathrm{Tr}(\underset{\sim}{H}\underset{\sim}{D}_4) \qquad (81)$$

$$a_6 = \tfrac{64}{27}\mathrm{Tr}(\underset{\sim}{H}\underset{\sim}{A}_6) \qquad (82)$$

$$b_6 = \tfrac{64}{5}\mathrm{Tr}(\underset{\sim}{H}\underset{\sim}{B}_6) \qquad (83)$$

$$c_6 = \tfrac{768}{35}\mathrm{Tr}(\underset{\sim}{H}\underset{\sim}{C}_6) . \qquad (84)$$

$\underset{\sim}{H}$ is the exact or the approximate effective Hamiltonian constructed from FCI or UHF solutions, respectively.

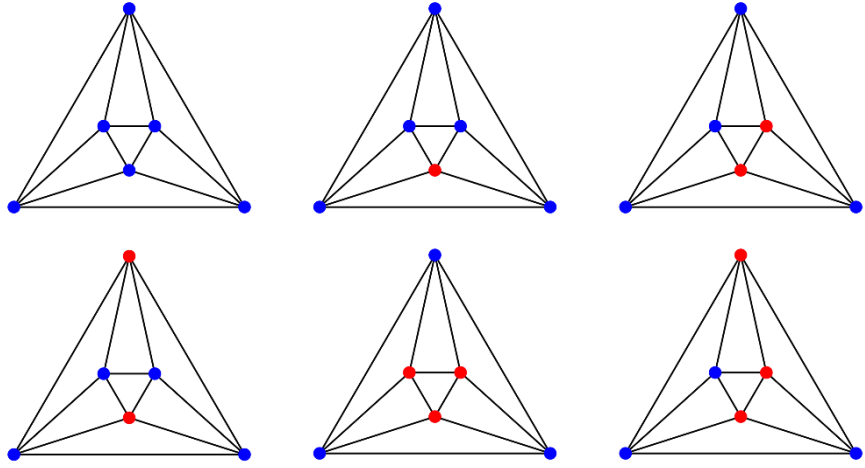

Figure 10: In the octahedron, there are six unique collinear spin configurations. Blue circles represent spin up, while red circles denote spin down. Beyond the high-spin (HS) configuration, there is one configuration with a single spin flipped, two with two flips, and two with three flips. Other configurations can be derived from these by site permutations according to the $O_h$ group, by inverting all spins, or a combination of both.

Based on energies only, six inequivalent configurations (Figure 10) would allow to extract five model parameters. Although one could potentially construct certain combinations of multi-center terms based on perturbation theory, such attempts would be unreliable in realistic scenarios, so we do not explore them here. Table 2 contrasts values obtained from exact calculations with those derived using the new BS method, demonstrating an overall good agreement in numerical values. Large relative deviations are only observed for less important parameters of smaller magnitude, specifically $J_{14}$, $a_6$ and $b_6$. All extracted values are to within numerical accuracy independent of whether we adapt to $\underset{\sim}{\mathbf{S}}^2$ symmetry. In Figure 11, we compare the eigenvalues for the exact and BS spin Hamiltonians, derived from the Hubbard model at $t = 0.15$ and $U = 1$. Each level is uniquely identified by an $O_h$ term symbol, and multiple degenerate substates comprising a term are illustrated.

Table 2: Comparison of spin-Hamiltonian parameters (cf. Eq. (75)) for the single-band Hubbard model on an octahedron with $t = 0.15$, $U = 1$.

|       | Exact     | BS       |
|-------|-----------|----------|
| $J_{12}$ | 0.07830  | 0.07807  |
| $J_{14}$ | 0.00083  | 0.00229  |
| $a_4$    | 0.14184  | 0.12985  |
| $b_4$    | -0.06054 | -0.06833 |
| $c_4$    | -0.14645 | -0.14361 |
| $d_4$    | 0.53358  | 0.52645  |
| $a_6$    | 0.02252  | 0.01629  |
| $b_6$    | -0.01612 | -0.00970 |
| $c_6$    | 0.07929  | 0.07616  |

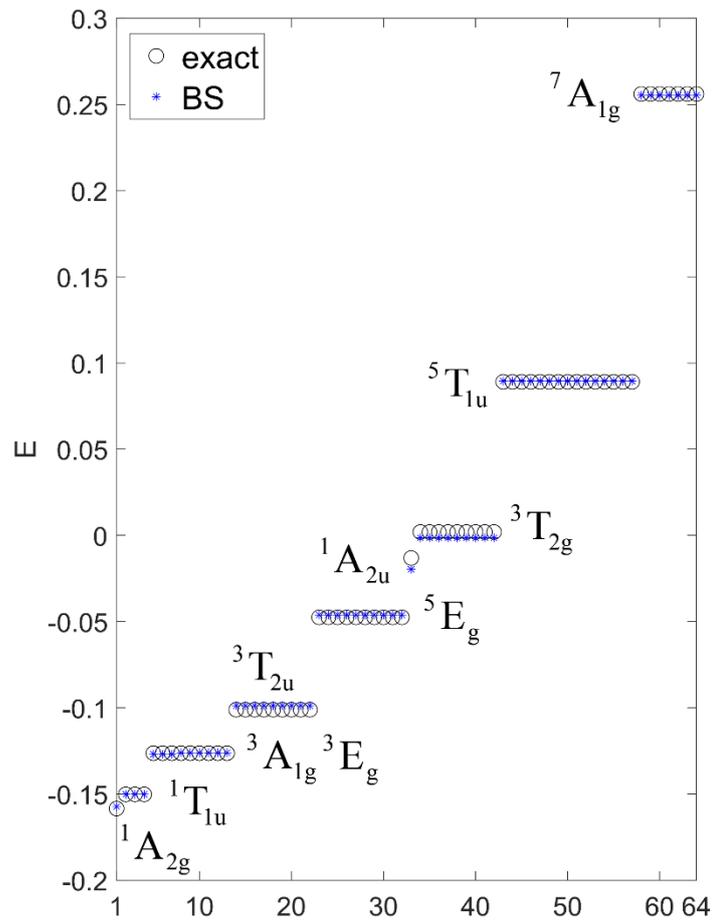

Figure 11: Illustration of the 64 eigenvalues for the isotropic $s = \frac{1}{2}$ Hamiltonian on the octahedron, derived from the Hubbard model with parameters $t = 0.15$ and $U = 1$. Exact eigenvalues, shifted to have a mean energy of zero (indicated as black spheres), are compared against those from the BS method (blue asterisks). Each eigenvalue level is labeled using $O_h$ term symbols. For degenerate levels, such as the $^3T_{2u}$ term, multiple degenerate substates are depicted.

The Heisenberg model on the octahedron can be solved analytically when written as Eq. (85),[59]

$$\underset{\sim}{H} = \tfrac{J}{2}(\underset{\sim}{\mathbf{S}}^2 - \underset{\sim}{\mathbf{s}}_{14}^2 - \underset{\sim}{\mathbf{s}}_{25}^2 - \underset{\sim}{\mathbf{s}}_{36}^2) \ , \tag{85}$$

which yields only five distinct energy levels for $s = \tfrac{1}{2}$, exhibiting accidental degeneracies between different terms. These degeneracies are resolved when incorporating 4c- and 6c-terms, underscoring the qualitative significance of multi-center exchange in this system. Upon exact diagonalization of the Hubbard Hamiltonian, we however observed an accidental degeneracy between $^3A_{1g}$ and $^3E_g$ terms (cf. Figure 11), whose cause we could not resolve. In any case, the Hubbard model can encompass additional symmetries not necessarily reflected in the spin Hamiltonian. This includes the pseudospin symmetry seen in bipartite systems[60,61] (though the octahedron does not represent a bipartite lattice). Moreover, highly non-trivial hidden dynamical symmetries might exist[62] which can induce accidental degeneracies, as exemplified by the six-site Hubbard ring investigated by Heilmann and Lieb.[60–62] Contrary to this, our spin Hamiltonian constructed from BS solutions does not manifest this accidental degeneracy; in its spectrum, the $^3A_{1g}$ term is marginally lower than the $^3E_g$ term (barely discernible from Figure 11).

## 4. Discussion

As recently emphasized,[17] accurately determining exchange parameters does not necessitate fitting to directly computed electronic eigenstates. The underlying principle of an effective model lies in the existence of a low-energy subspace. Within this subspace, where matrix elements of the model Hamiltonian and the *ab initio* Hamiltonian agree, any rotation can be chosen to define a basis that is convenient for computing the model parameters.[17] Fitting to eigenstates can be counterproductive if sufficiently accurate theoretical calculations are prohibitively demanding.[17] This underlines the value of BS methods. These circumvent the need to calculate exact electronic eigenstates and instead focus on capturing the essential spin interactions. This strategy proves particularly effective when dealing with multinuclear magnetic systems. From this perspective, categorizing BS states as unphysical[63] appears as an overly strict assessment.

In the context of the dimer and trimer, utilizing the effective-Hamiltonian technique to determine the exchange couplings might seem akin to using a sledgehammer to crack a nut.

This is because, for these simple systems, the energies from the generalized eigenvalue problem in the space of the BS solutions would be sufficient for this purpose. The true merit of the effective-Hamiltonian approach becomes evident when one cannot extract all parameters solely from the energy spectra.[29,31,36] While our new procedure can yield a complete set of spin-Hamiltonian parameters, this should not be mistaken to imply that all these parameters will always be significant in every case. The relevancy of certain parameters can greatly depend on the system under consideration. Besides, isotropic multi-exchange terms could be less critical in specific cases than anisotropic terms, which are not addressed here.

Note that cyclic exchange has been considered in interpreting experimental data (e.g., INS measurements[64,65]), and in theoretical[66–69] and quantum-chemical[29,70–72] studies, predominantly for cuprates. Six-spin interactions, on the other hand, could be relevant in clusters whose topology enables collective exchange processes, but to the best of our knowledge, they have to date not been explicitly considered for magnetic solids or molecules. Such terms are challenging to differentiate from other types of interactions, and their unambiguous identification would likely require advanced spectroscopic techniques. These obstacles do not impede our theoretical exploration of these terms. In fact, the ability to calculate and analyze higher-order interactions could guide experimental efforts by providing a theoretical understanding of the conditions under which 6c terms may have significant contributions.

This study does not address all issues that might be encountered in practice. As an example, consider a binuclear $s=1$ system, whose isotropic spin Hamiltonian comprises a Heisenberg and a biquadratic exchange term:[73]

$$\underline{H} = J(\underline{\mathbf{s}}_1 \cdot \underline{\mathbf{s}}_2) + j(\underline{\mathbf{s}}_1 \cdot \underline{\mathbf{s}}_2)(\underline{\mathbf{s}}_1 \cdot \underline{\mathbf{s}}_2) \,. \tag{86}$$

Associating HS $|\underline{\Uparrow\Uparrow}\rangle$ and BS $|\underline{\Uparrow\Downarrow}\rangle$ states with spin configurations $|m_1=+1, m_2=+1\rangle$ and $|m_1=+1, m_2=-1\rangle$, respectively, generates the underdetermined linear system of Eqs. (87) and (88),

$$\langle \underline{\Uparrow\Downarrow} | \hat{H} | \underline{\Uparrow\Downarrow} \rangle = -J + 2j + E_0 \,, \tag{87}$$

$$\langle \underline{\Uparrow\Uparrow} | \hat{H} | \underline{\Uparrow\Uparrow} \rangle = J + j + E_0 \,. \tag{88}$$

Typically, $j$ is neglected (set to zero) and the system is solved for $J$. A challenge that arises when attempting to compute a full set of BS states for systems with $s > \frac{1}{2}$ is that a single

Slater determinant cannot represent local spin projections that are not maximal in magnitude. For instance, while for an $s = 1$ site maximal and minimal projections, that is, $m = \pm 1$, can be adequately captured by a single Slater determinant, the intermediate state $m = 0$ does not lend itself to such a straightforward representation. Procedures to represent local $m = 0$ projections for the computation of biquadratic exchange in dinuclear $s = 1$ systems[30] are not entirely straightforward or broadly applicable. It is also worth noting that the Noodleman approach, when applied to non-collinear spin configurations, can provide further insight, e.g., for parametrizing anisotropic spin Hamiltonians,[40,74,75] or on biquadratic exchange.[72] However, this brings additional challenges related to defining local spin directions and imposing constraints to converge onto the desired solutions. In the context of the present strategy, information about biquadratic exchange can be gleaned from the off-diagonal matrix element between $|\Uparrow \Downarrow\rangle$ and its degenerate spin-flipped counterpart $|\Downarrow \Uparrow\rangle$:

$$\langle \Uparrow \Downarrow | \hat{H} | \Downarrow \Uparrow \rangle = j \ . \tag{89}$$

Despite this brief outline on the potential course of action for $s > \frac{1}{2}$, an exploration of such systems is beyond the scope of this study.

Another complication is the exponential growth in the number of spin configurations $2^N$, so the task of computing a BS state for each configuration rapidly becomes infeasible as $N$ increases. Consequently, for larger systems, it would be imperative to extract information about model parameters from a limited subset of BS states. One may still utilize not just energies, but also off-diagonal matrix elements, but an implementation of such a reduced scheme is left for future work.

Lastly note that the use of the UHF method is tailored to the single-band Hubbard model with an inherently reduced degree of dynamic correlation compared to most molecular electronic-structure problems. BS-HF is therefore suitable and efficient for this investigation. For more complex, realistic systems like magnetic molecules, capturing dynamical correlation becomes crucial to obtain quantitatively accurate results. Some options to incorporate dynamical correlation are many-body perturbation theory corrections, coupled-cluster theory, or DFT. In the latter case, DFT energies could be used on the diagonal of the Hamiltonian matrix in the basis of BS solutions, and off-diagonal elements might be calculated in terms of Kohn-Sham determinants. We believe that the utility of these perspectives warrants detailed investigation.

## 5. Summary and Conclusion

We have presented an extended broken-symmetry (BS) method that does not solely rely on energies, as energies might not always be adequate or sufficient in capturing all spin-Hamiltonian parameters in exchange-coupled clusters. For $s = \frac{1}{2}$ systems, we outlined a comprehensive framework for adapting the canonical effective-Hamiltonian theory to be based on BS solutions. A significant advantage of this method is its ability to evaluate all multi-center exchange terms in an unbiased fashion. In addition, we have briefly discussed the construction of linearly independent and totally symmetric sets of multi-exchange operators, to avoid ambiguities and redundancies when defining spin Hamiltonians.

The Hubbard model served as a testbed because it allows comparisons with exact-diagonalization results. In the Hubbard dimer, our approach is equivalent to the established Yamaguchi correction. Although our approach in its present form suffers from residual artifactual anisotropy, it was found to be slightly superior to the Yamaguchi correction in the triangle, especially as the $\frac{t}{U}$ ratio approaches the intermediate-exchange regime. In the Hubbard square, we noted the presence of two independent four-center exchange terms. Our method could quantify both, unlike the conventional energy-based Noodleman (N) procedure, which is restricted to choosing one. While the N mapping may yield acceptable results when the selected 4c-term is consistent with perturbation theory, relying on a basic single-band Hubbard model would not be reliable for all molecules or materials. Lastly, we successfully demonstrated that our method could determine all 2c, 4c, and 6c terms in the octahedron; the exact Hubbard spectrum was reproduced with appreciable accuracy at an intermediate $\frac{t}{U}$ ratio.

Our procedure can yield a comprehensive set of parameters for all possible multi-center terms, but this does not imply that every one of these is essential for understanding the primary magnetic properties of a system. Indeed, in many cases, a reduced set of key parameters captures the dominant features of the magnetic behavior. Unbiased quantum-chemical calculations can contribute to identifying these critical parameters and balancing between the complexity of the model and the fidelity to the physical reality. However, while our methodology offers a robust framework for understanding magnetic phenomena in model systems, there are obstacles to overcome before its practical application in realistic molecules or clusters can be realized. The absence of dynamic correlation in Hartree–Fock wave functions does not significantly impact the results for the single-band Hubbard model, but considering correlation effects will be indispensable for the application to realistic systems. Exploring these additional complications appears worthwhile to refine our generalized BS method and make it ready for practical use in magnetic clusters.

**Acknowledgement.** The author thanks Peter Saalfrank for generous support.

**Conflicts of interest.** The author has no conflicts of interest to disclose.

# References

[1] S. Blundell, *Magnetism in Condensed Matter* (Oxford University Press, 2001).

[2] A. Bencini, and D. Gatteschi, *Electron Paramagnetic Resonance of Exchange Coupled Systems* (Springer Berlin, 1990).

[3] D. Gatteschi, and R. Sessoli, Angew. Chemie Int. Ed. **42**, 268 (2003).

[4] J. Schnack, Contemp. Phys. **60**, 127 (2019).

[5] W. Marshall, and S.W. Lovesey, *Theory of Thermal Neutron Scattering* (Clarendon Press, Oxford, 1971).

[6] A. Furrer, and H.U. Güdel, Phys. Rev. Lett. **39**, 657 (1977).

[7] K.H. Marti, and M. Reiher, in *Prog. Phys. Chem. Vol. 3* (Oldenbourg Wissenschaftsverlag, 2011), pp. 293–309.

[8] G.K.-L. Chan, and S. Sharma, Annu. Rev. Phys. Chem. **62**, 465 (2011).

[9] J. Finley, P.-Å. Malmqvist, B.O. Roos, and L. Serrano-Andrés, Chem. Phys. Lett. **288**, 299 (1998).

[10] C. Angeli, R. Cimiraglia, S. Evangelisti, T. Leininger, and J.-P. Malrieu, J. Chem. Phys. **114**, 10252 (2001).

[11] M. Roemelt, V. Krewald, and D.A. Pantazis, J. Chem. Theory Comput. **14**, 166 (2018).

[12] M. Roemelt, and D.A. Pantazis, Adv. Theory Simulations **2**, 1800201 (2019).

[13] R. Maurice, C. De Graaf, and N. Guihery, Phys. Chem. Chem. Phys. **15**, 18784 (2013).

[14] W. Dobrautz, O. Weser, N.A. Bogdanov, A. Alavi, and G. Li Manni, J. Chem. Theory Comput. **17**, 5684 (2021).

[15] N.J. Mayhall, and M. Head-Gordon, J. Chem. Phys. **141**, (2014).

[16] N.J. Mayhall, and M. Head-Gordon, J. Phys. Chem. Lett. **6**, 1982 (2015).

[17] H. Schurkus, D.-T. Chen, H.-P. Cheng, G. Chan, and J. Stanton, J. Chem. Phys. **152**, 234115.

[18] L. Noodleman, J. Chem. Phys. **7**, 5737 (1981).

[19] L. Noodleman, and E.R. Davidson, Chem. Phys. **109**, 131 (1986).

[20] L. Noodleman, C.Y. Peng, D.A. Case, and J.M. Mouesca, Coord. Chem. Rev. **144**, 199 (1995).

[21] E. Ruiz, J. Cano, S. Alvarez, and P. Alemany, J. Comput. Chem. **20**, 1391 (1999).

[22] F. Illas, I.D.R. Moreira, C. de Graaf, and V. Barone, Theor. Chem. Acc. **104**, 265 (2000).

[23] E. Ruiz, A. Rodríguez-Fortea, J. Cano, S. Alvarez, and P. Alemany, J. Comput. Chem. **24**, 982 (2003).

[24] A. Bencini, and F. Totti, J. Chem. Theory Comput. **5**, 144 (2008).

[25] C. van Wüllen, J. Chem. Phys. **130**, 194109 (2009).

[26] K. Park, M.R. Pederson, and C.S. Hellberg, Phys. Rev. B **69**, 14416 (2004).

[27] S. Ghassemi Tabrizi, A. V Arbuznikov, and M. Kaupp, J. Phys. Chem. A **120**, 6864 (2016).

[28] H.F. Schurkus, D. Chen, M.J. O'Rourke, H.-P. Cheng, and G.K.-L. Chan, J. Phys. Chem. Lett. **11**, 3789 (2020).

[29] C.J. Calzado, and J.-P. Malrieu, Phys. Rev. B **69**, 94435 (2004).

[30] P. Labèguerie, C. Boilleau, R. Bastardis, N. Suaud, N. Guihéry, and J.-P. Malrieu, J. Chem. Phys. **129**, (2008).

[31] D. Reta, I. de P.R. Moreira, and F. Illas, J. Chem. Theory Comput. **12**, 3228 (2016).

[32] K. Yamaguchi, F. Jensen, A. Dorigo, and K.N. Houk, Chem. Phys. Lett. **149**, 537 (1988).

[33] M. Shoji, K. Koizumi, Y. Kitagawa, T. Kawakami, S. Yamanaka, M. Okumura, and K. Yamaguchi, Chem. Phys. Lett. **432**, 343 (2006).

[34] C. Bloch, Nucl. Phys. **6**, 329 (1958).

[35] J. des Cloizeaux, Nucl. Phys. **20**, 321 (1960).

[36] J.P. Malrieu, R. Caballol, C.J. Calzado, C. De Graaf, and N. Guihery, Chem. Rev. **114**, 429 (2013).

[37] To avert any confusion, operators acting in the original space (the electronic-structure problem) carry a caret, e.g., $\hat{H}$, while operators defined in the model space (the domain of the spin Hamiltonian) are indicated by a tilde underneath the symbol, e.g., $\utilde{H}$. The model-space projector $P_0$ does not bear either symbol.

[38] L.F. Chibotaru, Adv. Chem. Physics **153**, 397 (2013).

[39] S. Ghassemi Tabrizi, A. V Arbuznikov, and M. Kaupp, Chem. Eur. J. **24**, 4689 (2018).

[40] S. Ghassemi Tabrizi, A. V Arbuznikov, and M. Kaupp, J. Phys. Chem. A **123**, 2361 (2019).

[41] J.J. Sakurai, *Modern Quantum Mechanics*, 2nd ed., edited by San Fu Tuan (Addison Wesley, Reading, MA, 1993).

[42] In a minimal basis, high-spin states, e.g., $|\underline{\Uparrow\Uparrow}\rangle = |\Uparrow\Uparrow\rangle$, have well-defined total spin.

[43] C.A. Jiménez-Hoyos, Ph. D. Thesis, Rice University, Houston, TX, 2013.

[44] S. Ghassemi Tabrizi, and C.A. Jiménez-Hoyos, Phys. Rev. B **105**, 35147 (2022).

[45] S. Ghassemi Tabrizi, and C.A. Jiménez-Hoyos, Condens. Matter **8**, 18 (2023).

[46] A.H. MacDonald, S.M. Girvin, and D. Yoshioka, Phys. Rev. B **37**, 9753 (1988).

[47] J.P. Malrieu, and D. Maynau, J. Am. Chem. Soc. **104**, 3021 (1982).

[48] D. Maynau, and J.P. Malrieu, J. Am. Chem. Soc. **104**, 3029 (1982).

[49] Isotropic three-center terms exist in $s=1$ systems.

[50] K. Bärwinkel, H.-J. Schmidt, and J. Schnack, J. Magn. Magn. Mater. **212**, 240 (2000).

[51] O. Waldmann, Phys. Rev. B **61**, 6138 (2000).

[52] N.P. Konstantinidis, Phys. Rev. B **72**, 64453 (2005).

[53] A. Szabo, and N.S. Ostlund, *Modern Quantum Chemistry: Introduction to Advanced Electronic Structure Theory*, Dover Publications, Courier Corporation, 1996.

[54] N. Ferré, N. Guihéry, and J.-P. Malrieu, Phys. Chem. Chem. Phys. **17**, 14375 (2015).

[55] H. Fukutome, Int. J. Quantum Chem. **20**, 955 (1981).

[56] Calzado and Malrieu[29] quantified three distinct 4c parameters for square plaquettes derived from Cuprate lattices, apparently without realizing that there are only two independent 4c terms in an $s=\frac{1}{2}$ square. However, this does not make any of their results incorrect.

[57] $\utilde{P}_{1234}$ is a product of three pairwise exchanges, $\utilde{P}_{1234} = \utilde{P}_{12}\utilde{P}_{23}\utilde{P}_{34}$; for $s=\frac{1}{2}$, $\utilde{P}_{ij} = \frac{1}{2}(\utilde{1} + 4\utilde{\mathbf{s}}_i \cdot \utilde{\mathbf{s}}_j)$. $\utilde{P}_{1234}$ is the generator of the $C_4$ subgroup of $D_4$.

[58] A.A. Aligia, Phys. Rev. B **98**, 125118 (2018).

[59] J. Schnack, and M. Luban, Phys. Rev. B **63**, 14418 (2000).

[60] O.J. Heilmann, and E.H. Lieb, Ann. N. Y. Acad. Sci. **172**, 584 (1971).

[61] R. Schumann, Ann. Phys. **11**, 49 (2002).

[62] H. Grosse, Lett. Math. Phys. **18**, 151 (1989).

[63] S. Kotaru, S. Kähler, M. Alessio, and A.I. Krylov, J. Comput. Chem. **44**, 367 (2023).

[64] R. Coldea, S.M. Hayden, G. Aeppli, T.G. Perring, C.D. Frost, T.E. Mason, S.-W. Cheong, and Z. Fisk, Phys. Rev. Lett. **86**, 5377 (2001).

[65] S. Notbohm, P. Ribeiro, B. Lake, D.A. Tennant, K.P. Schmidt, G.S. Uhrig, C. Hess, R. Klingeler, G. Behr, B. Büchner, et al., Phys. Rev. Lett. **98**, 27403 (2007).

[66] M. Roger, and J.M. Delrieu, Phys. Rev. B **39**, 2299 (1989).

[67] Y. Honda, Y. Kuramoto, and T. Watanabe, Phys. Rev. B **47**, 11329 (1993).

[68] J. Lorenzana, J. Eroles, and S. Sorella, Phys. Rev. Lett. **83**, 5122 (1999).

[69] A. Läuchli, G. Schmid, and M. Troyer, Phys. Rev. B **67**, 100409 (2003).

[70] C.J. Calzado, C. de Graaf, E. Bordas, R. Caballol, and J.-P. Malrieu, Phys. Rev. B **67**(13), 132409 (2003).

[71] I. de PR Moreira, C.J. Calzado, J.-P. Malrieu, and F. Illas, New J. Phys. **9**, 369 (2007).

[72] N.S. Fedorova, C. Ederer, N.A. Spaldin, and A. Scaramucci, Phys. Rev. B **91**, 165122 (2015).

[73] Biquadratic exchange can be expressed as a linear combination of scalar couplings of rank-1 or rank-2 local spin operators. Although the rank-1 component could be incorporated into the Heisenberg term, we adhere to the convention of viewing biquadratic exchange as the square of rank-1 scalar couplings, see Eq. (86).

[74] R. Takeda, S. Yamanaka, M. Shoji, and K. Yamaguchi, Int. J. Quantum Chem. **107**, 1328 (2007).

[75] A. Wodýnski, and M. Kaupp, J. Chem. Theory Comput. **14**, 1267 (2018).